\AtBeginDocument{%
  \paperwidth=\dimexpr
    1in + \oddsidemargin
    + \textwidth
    + 1in + \oddsidemargin
  \relax
  \paperheight=\dimexpr
    1in + \topmargin
    + \headheight + \headsep
    + \textheight
    + 1in + \topmargin
  \relax
  \usepackage[pass]{geometry}\relax
}
\RequirePackage{fix-cm}
\documentclass[smallextended]{svjour3}       
\smartqed  
\usepackage{graphicx}
\usepackage{amsmath,amssymb,amsfonts}
\usepackage{tikz}
\usetikzlibrary{shapes.geometric, arrows}
\usepackage{algorithmic}
\usepackage{graphicx}
\usepackage{textcomp}
\usepackage{xcolor}
\usepackage{tcolorbox}
\usepackage{float}
\usepackage{adjustbox}
\usepackage{subfigure}
\usepackage{tabularx}
\usepackage{listings}
\usepackage{xcolor}
\usepackage{framed}
\colorlet{shadecolor}{gray!10}
\colorlet{framecolor}{black}
\usepackage[numbers]{natbib}
\usepackage[inline]{enumitem}
\usepackage{algorithmic}
\usepackage{graphicx,multirow}
\usepackage{subfigure}
\usepackage{hyperref}
\usepackage{adjustbox}
\usepackage{float}
\usepackage{makecell}
\usepackage{amsmath,amsfonts,amssymb}
\usepackage{textcomp}
\usepackage{xcolor}
\usepackage{enumitem}
\usepackage{balance}
\usepackage{hhline}
\usepackage{lscape}
\usepackage{rotating}
\usepackage{xcolor, soul}
\sethlcolor{green}
\usepackage[noframe]{showframe}
\usepackage{framed}
\usepackage{lipsum}
\definecolor{lightgray}{gray}{0.75}
\usepackage{graphicx}
\usepackage{subcaption}

{\endMakeFramed}
\usepackage{booktabs}

\def\BibTeX{{\rm B\kern-.05em{\sc i\kern-.025em b}\kern-.08em
    T\kern-.1667em\lower.7ex\hbox{E}\kern-.125emX}}

\begin{document}

\title{Towards Enhancing the Reproducibility of Deep Learning Bugs: An Empirical Study}

\author{Mehil B. Shah         \and
         Mohammad Masudur Rahman \and 
         Foutse Khomh
}

\institute{Mehil Shah \at
              Dalhousie University, Canada \\
              \email{shahmehil@dal.ca}
           \and
           Mohammad Masudur Rahman \at
           Dalhousie University, Canada \\
           \email{masud.rahman@dal.ca}
           \and
           Foutse Khomh \at
           Polytechnique Montreal, Canada \\
           \email{foutse.khomh@polymtl.ca}
}

\date{Received: date / Accepted: date}

\maketitle

\begin{abstract}
\hspace{0pt}\vspace{0.2\baselineskip}

\textbf{Context:} Deep learning has achieved remarkable progress in various domains. However, like any software system, deep learning systems contain bugs, some of which can have severe impacts, as evidenced by crashes involving autonomous vehicles. Despite substantial advancements in deep learning techniques, little research has focused on reproducing deep learning bugs, which is an essential step for their resolution. Existing literature suggests that only 3\% of deep learning bugs are reproducible, underscoring the need for further research.

\textbf{Objective:} This paper examines the reproducibility of deep learning bugs. We identify edit actions and useful information that could improve the reproducibility of deep learning bugs.

\textbf{Method:} First, we construct a dataset of 668 deep learning bugs from Stack Overflow and GitHub across three frameworks and 22 architectures. Second, out of the 668 bugs, we select 165 bugs using stratified sampling and attempt to determine their reproducibility. While reproducing these bugs, we identify edit actions and useful information for their reproduction. Third, we used the Apriori algorithm to identify useful information and edit actions required to reproduce specific types of bugs. Finally, we conduct a user study involving 22 developers to assess the effectiveness of our findings in real-life settings. 

\textbf{Results:} We successfully reproduced 148 out of 165 bugs attempted. We identified ten edit actions and five useful types of component information that can help us reproduce the deep learning bugs. With the help of our findings, the developers were able to reproduce 22.92\% more bugs and reduce their reproduction time by 24.35\%.

\textbf{Conclusions:} Our research addresses the critical issue of deep learning bug reproducibility. Practitioners and researchers can leverage our findings to improve deep learning bug reproducibility.
\end{abstract}

\section{Introduction}
Deep learning (hereby DL) has been widely used in many application domains, including natural language processing, finance, cybersecurity~\cite{med1, fin1, cyb1}, autonomous vehicles~\cite{pei2017, ma2018}, and healthcare systems~\cite{esteva2019guide}.

Like any software system, deep learning systems are prone to bugs. Bugs in deep learning systems could arise from various sources, including errors in the dataset, incorrect hyperparameters, and incorrect structure of the deep learning model~\cite{BRAIEK2020110542}. These bugs could lead to program failures, poor performance, or incorrect functionality, as reported by existing literature~\cite {islamfse19}. They can also lead to serious consequences, as shown by the fatal crash involving Uber's self-driving car~\cite{selfdrivingcarcrash}. Thus, we must fix the bugs before deploying a deep learning model in production. However, one must reproduce the bugs before fixing them to verify their presence/absence in the system. Unfortunately, reproducing DL bugs is challenging due to deep learning systems' multifaceted nature and dependencies, which encompass data, hardware, libraries, frameworks, and client programs. Furthermore, deep learning systems are inherently non-deterministic (i.e., random weight initialization), which leads to different outcomes across multiple runs and thus makes the reproduction of deep learning bugs challenging~\cite{nagarajan2018impact}. Moreover, they also suffer from a lack of interpretability~\cite{krishnan2020against}, which makes the reproduction of deep learning bugs challenging in comparison to traditional software bugs. According to existing investigations~\cite{defects4ml}, only 3\% of their analyzed deep learning bugs were reproducible, further demonstrating the challenges in reproducing DL bugs.

Existing literature investigates the challenges of reproducing programming errors or bugs from various sources. Mondal et al.~\cite{msr2019mondal} investigate the challenges in reproducing programming issues reported on Stack Overflow and suggest several edit actions to help reproduce them. Rahman et al.~\cite{icsme2020rahman} conduct a multi-modal study to understand the factors behind the non-reprodu\-cibility of software bugs. They identify 11 significant factors behind bug non-reproducibility, including missing information, bug duplication, false positive bugs, and intermittency. Overall, these studies highlight the challenges in reproducing traditional software bugs but do not deal with any deep learning bugs, which warrants further investigation.

Recently, a few studies have attempted to tackle the challenges of deep learning bugs. Liang et al.~\cite{liang2022gdefects4dl} provide a dataset of 64 deep learning bugs collected from GitHub issues. They classify these bugs into six categories according to the taxonomy of Humbatova et al.~\cite{taxonomyRealFaults}. Moravati et al.~\cite{defects4ml} constructed a benchmark dataset containing 100 deep learning bugs collected from StackOverflow and GitHub; they reproduced each of them. Although these studies offer benchmark datasets containing reproducible bugs, their primary focus is dataset construction. They do not report any detailed instructions (e.g., sequence of actions) essential to the reproduction of the deep learning bugs. Our work attempts to fill this important gap in the literature. 

In this paper, we conduct an empirical study to better understand the challenges in reproducing deep learning bugs. First, we collect a total of 568 DL bugs from Stack Overflow posts. Then, we extend them with 100 DL bugs from the benchmark dataset of Moravati et al.~\cite{defects4ml}, which makes up our final dataset of 668 DL bugs. Using the taxonomy of Humbatova et al.~\cite{defects4ml}, we divide these bugs into five categories: 167 model, 213 tensor, 145 training, 113 GPU, and 30 API bugs. Second, by using stratified sampling, we select 165 of these bugs and determine their reproducibility status by attempting to reproduce them using the code snippet and complementary information from Stack Overflow and the benchmark dataset. Third, the first author manually analyzed the produced artefacts to determine the categories of information that are useful for deep learning bug reproduction. Fourth, we use the information gathered during the bug reproduction and create a dataset of transactions that link the type of bug with edit actions and component information. Then, our study establishes connections among the component information, edit actions, and the type of bugs by employing the Apriori algorithm~\cite{apriori}. Finally, to further validate our findings, we conducted a developer study with 22 participants from industry and academia. Half of the participants were asked to reproduce bugs using our identified information, while the other half were asked to reproduce the same bugs without access to that information. Results from the user study show that our recommended edit actions and component information can reduce the time to reproduce the bug by 24.35\%. Thus, we answer three important research questions as follows: 
\begin{itemize}
\item \textbf{RQ1: Which edit actions are useful for reproducing deep learning bugs?
}\\
Determining the key edit actions that are crucial for reproducing deep learning bugs is important since almost none of the bugs can be reproduced using the verbatim code snippet. By manually reproducing 148 deep learning bugs, we identify ten key edit actions that could be useful to reproduce deep learning bugs (e.g., input data generation, neural network construction, hyperparameter initialization).
\item \textbf{RQ2: What types of component information and edit actions are useful for reproducing specific types of deep learning bugs?
}\\
Different types of DL bugs need different types of information or edit actions, which warrants further investigation. Using the Apriori algorithm, we have determined the top 3 pieces of information and the top 5 edit actions that can help one reproduce each type of bug. These insights can be used not only to detect the missing information in a submitted bug report but also to formulate the follow-up questions soliciting the missing information.
\item \textbf{RQ3: How do the suggested edit actions and information affect the reproducibility of deep learning bugs?
}\\
To assess the effectiveness of our suggested edit actions and component information in improving bug reproducibility, we conducted a user study involving 22 professional developers. In our developer study, the participants assigned to the experimental group used our suggested edit actions and component information to reproduce deep learning bugs. In contrast, the participants in the control group reproduced the bugs without access to the suggested edit actions and component information. We found that our recommended edit actions and component information (a) helped the developers reproduce 22.92\% more bugs and (b) decreased their time to reproduce the deep learning bugs by 24.35\%.
\end{itemize}

The remainder of the paper is organized as follows. Section \ref{section2} presents a motivating example that highlights the challenges in reproducing deep learning bugs. Section~\ref{section3} introduces the necessary background information for deep learning bugs and their bug reports. Section~\ref{section4} then describes our study methodology in detail, including data collection, dataset construction, environment setup, qualitative analysis, and user study design. Next, Section~\ref{section5} presents the findings by answering our three research questions. It summarizes the key edit actions, component information, and their relationships with bug types that enhance deep learning bug reproducibility. Section~\ref{section6} discusses our results and shows how our findings can be used to improve LLMs. Section~\ref{section7} discusses the threats to the validity of this study. Finally, Section~\ref{section8} reviews related literature, and Section~\ref{section9} concludes the paper.

\begin{figure}[h]
    \centering
\subfigure[Bug reported by the Stack Overflow User]{\label{fig:irreproducibleMotivatingExample}\includegraphics[width=0.8\columnwidth]{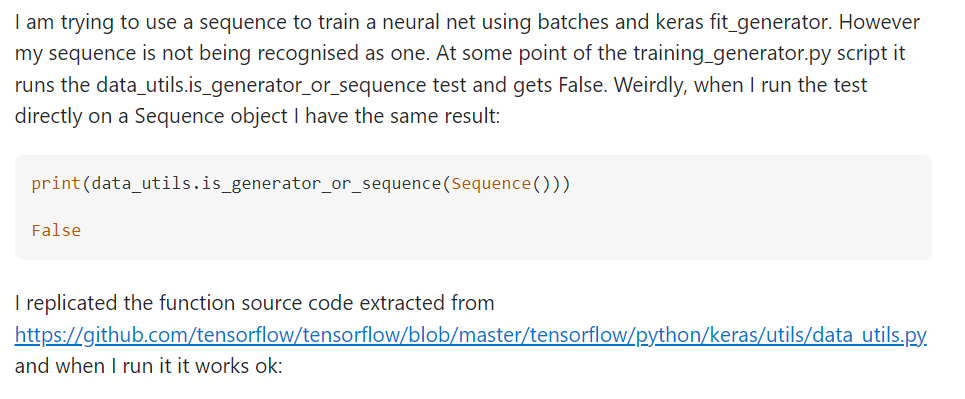}}
    \centering
    \hfill
    \subfigure[Answers highlighting the non-reproducibility of the bug]{\label{fig:nonReproducibleComments}\includegraphics[width=0.8\columnwidth]{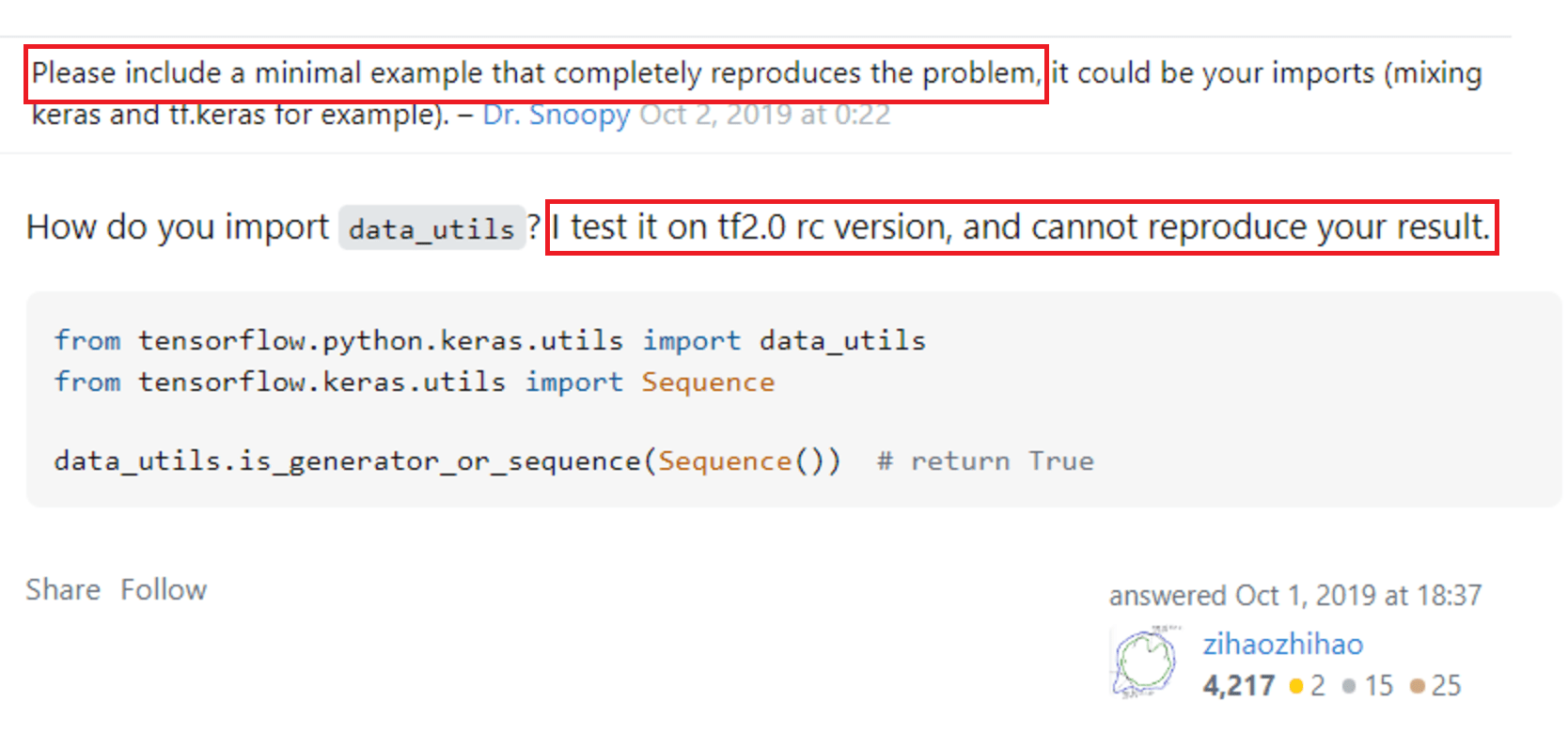}}
\caption{\textbf{An Irreproducible Bug} from Stack Overflow}
\end{figure}

\section {Motivating Example} \label{section2}
Bug reports or programming Q\&A posts might not always provide sufficient information to reproduce a deep learning bug. Let us consider the example question shown in Fig.~\ref{fig:irreproducibleMotivatingExample}~\cite{motivatingExample1}. Here, the user attempts to train a neural network using Keras. The input data type is a $Sequence$ object, which is used as the base object for fitting a sequence of data, such as a dataset~\cite{sequence} The user aims to pass the training dataset as a $Sequence$ object to the $fit\_generator()$ method. However, s/he discovers that their $Sequence$ object is not recognized by the Keras library. The user also provides a code snippet to aid the reproducibility of the bug. Unfortunately, the issue cannot be reproduced since the provided information does not contain the required dependencies and imports. In Stack Overflow, this question has failed to receive a precise response. Even though the above code (Fig.~\ref{fig:irreproducibleMotivatingExample}) could be made compilable or executable using edit actions, the bug cannot be reproduced due to its complex nature. Since the earlier study~\cite{msr2019mondal} does not deal with any deep learning bugs, their suggested edits might also not be effective.

\begin{figure*}
    \centering
\subfigure[Bug reported by the Stack Overflow User]{\includegraphics[width=0.8\columnwidth]{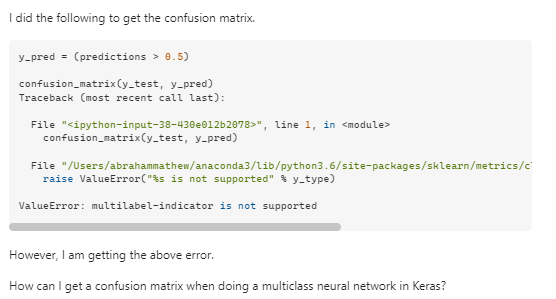}\label{fig:motivatingExample}}
    \centering
    \hfill
    \subfigure[Accepted Answer]{\label{fig:AcceptedAnswer}\includegraphics[width=0.8\columnwidth]{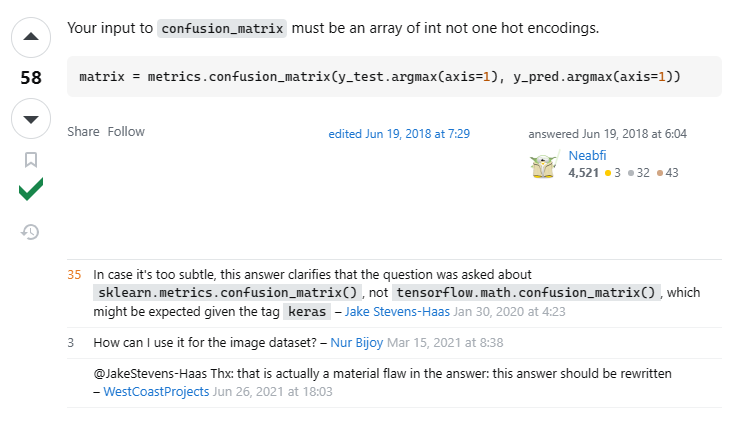}}
\caption{\textbf{A Reproducible Bug} from Stack Overflow}
\end{figure*}

Let us consider another example question shown in Fig.~\ref{fig:motivatingExample}~\cite{motivatingExample2}. Here, the user wants to obtain the confusion matrix from a multi-class classification model. Unfortunately, s/he runs into a runtime error, as the $confusion$ $\_matrix()$ method does not support multiple output labels. To reproduce this bug, we first generate a synthetic multi-class dataset since the training data is missing from the question. We also add the required import statement for the `confusion\_matrix' function. Then, we initialize the model hyperparameters based on the code snippet. When we run the code on this synthetic data, it triggers the same runtime error due to the wrong data shape being passed to `confusion\_matrix()'. Thus, by extracting the key information and applying the necessary edit actions, we were able to reproduce the bug. Similarly, this issue was reproduced by other users of Stack Overflow, and the question received a correct solution within two hours of its submission, as shown in Fig.~\ref{fig:AcceptedAnswer}. To summarize, by generating a synthetic dataset, adding the necessary import statement, and initializing the model hyperparameters, we can reproduce the example deep learning bug. Our study in this article proposes the methodology to extract the information and determine the edit actions necessary for systematically reproducing deep learning bugs.

\section{Background} \label{section3}
\subsection{\textit{Deep Learning Bugs}}
Developers often encounter different types of bugs when they write their deep-learning software. According to existing literature~\cite{taxonomyRealFaults, islam2019fse}, these bugs can be classified into five categories: Training, Model, Tensor and Input, API, and GPU. Table.~\ref{tab:prevalance} shows the prevalence of these bugs in deep learning systems. We discuss these bugs in detail below.
\begin{itemize}
\item \textbf{Training Bugs}: Training bugs encompass a wide range of issues that impact various aspects of the training process, including data preprocessing, hyperparameter tuning, loss function selection, and model optimization. Data preprocessing steps, such as normalization and input scaling, are frequently overlooked or improperly executed, leading to subpar results. Quality of training data is another critical factor that can affect of the training process~\cite{croft2023data}. Insufficient, imbalanced, or incorrectly labelled training data can significantly impact the model's ability to generalize and produce accurate predictions. Hyperparameter problems are quite common, with issues such as incorrect learning rates, inappropriate batch sizes, and inadequate selection of the number of epochs~\cite{taxonomyRealFaults}. Similarly, the selection and implementation of loss functions could be suboptimal or flawed. The choice and tuning of optimizers can also pose challenges, as incorrect selections or settings can hinder the model performance. Additional problems that may arise during training include memory mismanagement and the absence of data augmentation techniques. Overall, the multifaceted nature of neural network training, with its numerous hyperparameters and intricate architectures, often introduces bugs into the training process.
\item \textbf{Model Bugs}: Model bugs stem from the neural network architectures and their components. Common faults include wrong model type for a task, incorrect network depth with too few or too many layers, and issues with specific layers. For instance, choosing a multilayer perceptron over a convolutional neural network for an image classification task leads to poor performance, as the model type does not fit the problem. Similarly, networks with inadequate or excessive depth may fail to capture patterns or may overfit, respectively. Faulty layer design, such as the omission of data manipulation layers, could lead to flaws in feature extraction, whereas redundant layers increase the complexity of the model. Other frequent sources of bugs are suboptimal layer parameters like neuron counts, kernel dimensions, and stride lengths. Finally, inaccurate activation function selection, such as incorrectly using sigmoid instead of ReLU in the hidden layers, can degrade the model performance. Overall, model bugs stem from improper network design and configuration, leading to a fundamentally unsuitable model structure.
\begin{table}[]
\centering
\caption{Prevalence of Different Types of Deep Learning Bugs}
\label{tab:prevalance}
\resizebox{\columnwidth}{!}{%
\begin{tabular}{|l|l|l|l|l|l|}
\hline
\textbf{Type of Bug} & \textbf{Training}           & \textbf{Model}              & \textbf{Tensor and Input}   & \textbf{API}               & \textbf{GPU}               \\ \hline
\textbf{Prevalence}  & \multicolumn{1}{c|}{52.5\%} & \multicolumn{1}{c|}{19.7\%} & \multicolumn{1}{c|}{19.5\%} & \multicolumn{1}{c|}{5.3\%} & \multicolumn{1}{c|}{2.9\%} \\ \hline
\end{tabular}%
}
\end{table}
\item \textbf{Tensor \& Input Bugs}: Tensor bugs stem from incorrect tensor shapes provided in the deep learning models, i.e. mismatches between the expected tensor dimensions and the actual shape of the input tensor. Potential causes include errors in padding, indexing, or tensor transformations like transposing before model ingestion. Input bugs occur when the input data does not match the specifications the model requires. Examples include incorrect data types (e.g. strings instead of numeric), inconsistent data shapes (e.g. wrong matrix dimensions), and improper data formats (e.g. channels reversed in images). While tensor bugs arise from incorrect shaping of the input data, input bugs stem from invalid or incompatible characteristics of the raw input itself. Both categories can seriously hurt the model performance by propagating flawed representations, even if no crashes occur.

\item \textbf{API Bugs}: API bugs primarily occur due to an incorrect usage of the API. Several factors can lead to incorrect API usage. First, API definitions often change across framework versions, causing developers to use outdated or incompatible syntax. Second, there can be API incompatibility within the same framework. For example, in TensorFlow 1.x, the developers could use the Estimator API with \textit{model\_fn()} to build models. But, in TensorFlow 2.0, Estimators were removed, so code written for v1 would cause compile-time errors when the framework was updated. Finally, unclear or confusing documentation can result in developers misunderstanding the proper use of an API. This can lead to multiple issues, such as using an API incorrectly in a manner that deviates from the framework's intended use. Furthermore, developers may encounter challenges when the documentation lacks clear specifications of crucial API calls for a workflow or misplaces API call explanations within the code.
\item \textbf{GPU Bugs}: GPU bugs are the most difficult bugs to localize and reproduce, as observed by Jahan et al.~\cite{jahan2024towards}. GPU bugs can occur due to various factors such as incorrect GPU device referencing, failed parallelism, improper state sharing between subprocesses, and faulty data transfers to the GPU. Incorrect device referencing stems from developers misunderstanding the mapping between CPU and GPU environments~\cite{sanders2010cuda}. Failed parallelism and improper state sharing often result from subtle data races or synchronisation errors, respectively~\cite{aviram2012efficient, zhang2018g}. Faulty data transfers arise from mismatches between CPU and GPU data formats~\cite{tiwari2015understanding}. Overall, the domain-specific nature of GPU programming, the rapid evolution of GPU architectures, and the lack of sufficient validation could lead to difficult-to-diagnose GPU bugs.
\end{itemize}

\subsection{\textit{Deep Learning Bug Reports}}
To date, only limited investigation has been conducted to understand the patterns of deep learning (DL) bug reports. Long et al.~\cite{long2022reporting} performed the first exploratory study to analyze the bug reporting trends and patterns for deep learning frameworks. Their work revealed several key themes that provide insight into the nature and lifecycle of these bug reports:

\textbf{Root Causes and Prevalence:} Their study has found that low training speed is the most common symptom for submitting performance-related bug reports, ranging from 27\% to 67\% across the different DL frameworks. However, no consistent pattern was observed for the root causes of accuracy-related reports. This suggests that performance issues, especially those manifesting as slow execution speed, are a major pain point encountered by the users of DL frameworks.

\textbf{Affected Stages of DL Pipeline:} Across the frameworks studied, the training stage was found to be the most prevalent in performance and accuracy bug reports, ranging from 38\% to 77\%. This finding is not surprising, as training is typically the most computationally intensive and time-consuming stage of the DL pipeline~\cite{shi2016benchmarking}. Performance bottlenecks or accuracy issues encountered during training can significantly impact the overall usability and efficiency of the framework. 

\textbf{Report Quality and Resolution:} Their study found that a majority of the closed reports (69\% to 100\%) were either not classified, or their titles, labels, and content did not match the actual bugs reported. Furthermore, around 50\% of the reports that did reveal bugs were not resolved by direct code patches. These findings highlight inefficiencies in the bug reporting and resolution process.

In summary, the paper classifies performance and accuracy bug reports based on symptoms, pipeline stages affected, and GitHub resolution states to better understand the bugs from deep learning frameworks.

\section{Study Methodology} \label{section4}
Fig.~\ref{fig:schematicDiagram} shows the schematic diagram of our empirical study. We discuss different steps of our study as follows.
\subsection{\textit{Selection of Data Sources}}
\begin{figure*}[]
\centering
\adjustbox{fbox}{\includegraphics[width=\columnwidth]{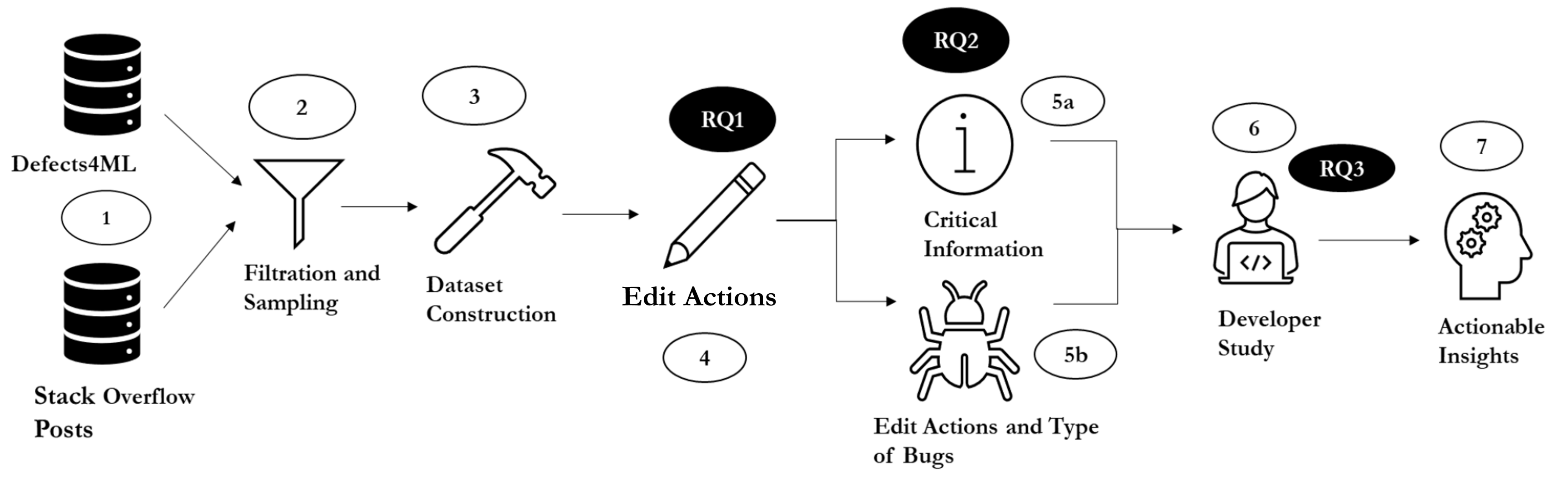}}
\caption{Schematic diagram of our empirical study}
\label{fig:schematicDiagram}
\end{figure*}

We select Stack Overflow as a primary data source for our study. It is the largest programming Q\&A site for programming topics containing over 24 million questions and 35 million answers~\cite{SOUsageData}. Thus, Stack Overflow could be a potential source for deep learning (DL) bugs. Developers often submit their encountered problems on Stack Overflow, when building their deep learning applications. According to a recent work~\cite{zhao2021state}, Stack Overflow contains at least 30 topics and 80K posts related to deep learning issues, which makes it a potential source for our data. We also select the benchmark dataset of Moravati et al.~\cite{defects4ml} (Defects4ML in Fig.~\ref{fig:schematicDiagram}) containing 100 DL bugs as our second source of data. Specifically, we use 63 GitHub issues and 37 Stack Overflow posts from the Defects4ML dataset in our study.

To gather relevant posts from Stack Overflow, we use the Stack Exchange Data Explorer platform\footnote{https://data.stackexchange.com/stackoverflow/query/new}.
We employ multiple filters to capture relevant posts discussing deep learning bugs. First, we select the recent posts (i.e., submitted between May 2020 and May 2023) with the following tags: `tensorflow', `keras', and `pytorch'. These tags were selected since they represent the most frequently used frameworks for deep learning~\cite{zhang2019empirical}. This filtration resulted in 14,065 posts for PyTorch, 14,971 posts for Tensorflow, and 3,152 posts for Keras. We then applied the following four filtration criteria to remove conceptual questions from our collection:
\begin{itemize}
    \item \textbf{Keyword Filtering}: This filtration removes the \textit{how-to} questions and the questions requesting installation instructions. To filter these questions, we use appropriate keywords (`how', `install', and `build'), as recommended by Humbatova et al.~\cite{taxonomyRealFaults}.
    \item \textbf{Code Snippet}: Given our focus on the reproducibility of reported issues, the questions must include code segments. Therefore, we consider such questions that contain at least one line of code, as suggested by Mondal et al.~\cite{msr2019mondal}.
    \item \textbf{Accepted Answer}: We only select the questions with accepted answers, ensuring that each reported issue (e.g., DL bug) was reproduced and fixed, as advised by Humbatova et al.~\cite{taxonomyRealFaults}.
    \item \textbf{Negative Question Score}: This filtration helps us discard questions that the community has found to be of low quality, as proposed by Ponzanelli et al.~\cite{ponzanelli2014}.
\end{itemize}
We combine all four filtration criteria above, construct a SQL query, and then execute the query against the Stack Exchange Data Explorer~\cite{StackExchange}. The SQL query is available in our replication package~\cite{replicationPackage}. After this filtration step, we obtained a total of 279 posts for Keras, 1,433 posts for Tensorflow, and 1,700 posts for PyTorch. To ensure the validity of our chosen filtration criteria and confirm that the posts are related to DL bugs, we conducted a follow-up manual analysis on a representative sample of the filtered posts. We determined the appropriate sample size using Cochran's sample size formula~\cite{cochran1977sampling}:

\begin{equation*}
n = \frac{Z^2 \cdot p \cdot (1-p)}{e^2}
\end{equation*}

Where:
\begin{itemize}
\item $n$ = sample size
\item $Z$ = Z-value for the desired confidence level (1.645 for 90\% confidence)
\item $p$ = population proportion (assumed to be 0.5 for maximum variability)
\item $e$ = margin of error (0.1 or 10\%)
\end{itemize}

After plugging in the values:

\begin{equation*}
n = \frac{1.645^2 \cdot 0.5 \cdot (1-0.5)}{0.1^2} = 67.65
\end{equation*}

Therefore, a representative sample of 67 posts was needed to achieve 90\% confidence with a 10\% margin of error. We chose these parameters to balance the precision of the estimates with the manual effort required for the analysis. Moreover, higher confidence levels or lower margins of error would have necessitated significantly larger sample sizes and analysis time as follows:

\begin{itemize}
\item 95\% confidence, 5\% margin of error: $n = 346$ posts (86.5 hours)
\item 99\% confidence, 1\% margin of error: $n = 2,832$ posts (708 hours)
\end{itemize}

For each of the 67 sampled posts, we analyzed the entire Q\&A discussion thread from Stack Overflow and the corresponding code changes in GitHub. This manual analysis took approximately 15 minutes per post, resulting in a total of around 17 hours of effort. We found that $\approx$96\% of the sampled posts discussed deep learning bugs, confirming the validity of our filtration criteria (Steps 1-2, Fig.~\ref{fig:schematicDiagram}).

\subsection{\textit{Dataset Construction}}
The above filtration step resulted in 3,412 Stack Overflow posts. To construct our final dataset, we apply another filter to these posts where we leverage the tags from Humbatova et al.\cite{taxonomyRealFaults}. We aimed to select representative samples from different types of deep learning bugs. To reduce the risk of miscategorizing bug types, we selected tags from Table 1 that exactly matched the leaves of the bug taxonomy proposed by Humbatova et al.~\cite{defects4ml}. We identified several Stack Overflow tags that partially matched the leaves of the taxonomy. However, to eliminate confusion about bug types, we only used tags that had an exact match with the taxonomy. Through this manual analysis of posts, bugs and tags, we selected the tags that best represented each bug taxonomy/sub-taxonomy from Humbatova et al.~\cite{taxonomyRealFaults}. Table.~\ref{tab:tagFiltration} shows our selected 3 tags for model bugs, 1 for tensor bugs, 11 for training bugs, 3 for GPU bugs, and 7 for API bugs. We look for these tags in the above Stack Overflow posts and collect 113 Model bugs, 193 Tensor Bugs, 95 Training Bugs, 101 GPU Bugs, and 24 API Bugs. We also discover 42 bugs belonging to multiple categories (e.g., 5 Model \& Tensor Bugs, 5 Tensor \& GPU Bugs, 3 Training \& API Bugs). Thus, the final dataset contains a total of 668 bugs (568 from our dataset + 100 from Defects4ML~\cite{defects4ml}) and captures a balanced representation from different types of DL bugs, as shown in Table~\ref{tab:datasetStatistics} (Step 3, Fig.~\ref{fig:schematicDiagram}).
\begin{table}
\centering
\captionsetup{justification=centering,singlelinecheck=false}
\caption{Tags used for filtering different types of bugs}
\resizebox{0.8\columnwidth}{!}{%
\begin{tabular}{|l|l|}
\hline
\textbf{Type of Bug} & \textbf{Tags} \\ \hline
Model & layer, model, activation-function \\ \hline
Tensor & tensor \\ \hline
Training & loss-function, training-data, optimization, loss, \\ & data-augmentation, performance, learning-rate, \\ & hyperparameters, initialization, imbalanced-data, nan \\ \hline
GPU & gpu, nvidia, cuda \\ \hline
API & typeerror, valueerror, attributeerror, importerror, \\ & compilererrors, syntaxerror, modulenotfounderror \\ \hline
\end{tabular}%
}
\label{tab:tagFiltration}
\end{table}

\begin{table}[]
\centering
\captionsetup{justification=centering,singlelinecheck=false} 
\caption{Summary of the constructed dataset}
\label{tab:datasetStatistics}
\resizebox{0.5\columnwidth}{!}{%
\begin{tabular}{|c|c|c|}
\hline
\textbf{Type of Bugs} & \textbf{Total Number of Bugs}  \\ \hline
Model          & 113          \\ \hline
Training       & 95           \\ \hline
GPU            & 101          \\ \hline
API            & 24           \\ \hline
Tensor and Input & 193  \\ \hline
Mixed          & 42            \\ \hline
\textbf{Total} & \textbf{568} \\ \hline
\end{tabular}%
}
\end{table}
\subsection{\textit{Environment Setup}}
For our experiments, we use the following environment setup:
\begin{itemize}
	\item \textbf{Code Editors}: We use \textit{Visual Studio Code v1.79.0}\footnote{https://code.visualstudio.com/} and \textit{PyCharm 2023.1.1}~\footnote{https://www.jetbrains.com/pycharm/} to execute code snippets and reproduce bugs from our dataset. Visual Studio Code and PyCharm are popular code editors for building DL-based applications~\cite{GeeksforGeeks_ide_dl}.
	\item \textbf{Dependencies}: To detect the API libraries adopted by the code snippets, we use the \textit{pipreqs} package\footnote{https://pypi.org/project/pipreqs/}. We also install the dependencies for each bug into a separate virtual environment using the \textit{venv}\footnote{https://docs.python.org/3/library/venv.html} module. 
	\item \textbf{Frameworks}: Since our dataset contains bugs from \textit{Tensorflow}, \textit{Keras}, and \textit{PyTorch}, we used all three frameworks in our experiments.
	\item \textbf{Libraries}: For generating the random inputs and visualizing the training metrics, we leveraged several scientific computing libraries such as \textit{numpy}, \textit{pandas}, and \textit{matplotlib}.
	\item \textbf{Python Version}: During our experimentation, we used \textit{Python v3.10} to reproduce the deep learning bugs, the latest stable version. If the Stack Overflow issue reported a Python version other than v3.10, we reproduce the bug using the mentioned version instead.
	\item \textbf{Hardware Config}: Our experiments were run on a desktop computer having a 64-bit Windows 11 Operating System with 16GB primary memory (i.e., RAM) and 8GB GPU Memory (Intel(R) Iris XE Graphics).
\end{itemize}
\subsection{\textit{Qualitative Analysis}}
Before conducting any qualitative analysis, we first determine if each post targets a deep learning bug. If a post is not related to a deep learning bug, we exclude it from further analysis. We also excluded the posts not reporting the evaluation metrics of the buggy model from our study. Without these metrics, it was impossible to determine if a bug was successfully reproduced or not. This two-step filtering process helped our analysis focus on relevant and measurable deep-learning bug reports. 

Once we confirm that the post is relevant to a DL bug, we check if each selected post accurately represents the type of bug it is assigned to. To validate the bug categorization, we manually analyze each selected Stack Overflow post and review the issue description, stack trace, observed behaviour, expected behaviour, and accepted answer. This ensures that the tag-based selection of posts from Stack Overflow does not affect our results' validity. We perform this manual verification since the Stack Overflow posts might lack an independently verified category label. In contrast, the GitHub issues in the benchmark dataset by Morovati et al.~\cite{defects4ml} already contain a validated category label. 

\subsubsection{Reproducing Deep Learning Bugs from Stack Overflow}
Once we confirm the category of each post or issue, we follow a two-step approach to reproduce the deep-learning bugs from Stack Overflow posts. First, we gather complementary information about each bug, such as the dataset, code snippet, library versions, and the framework used. Second, we attempt to reproduce the bug using the code snippet and supporting data (e.g., dataset information, environment configurations, hyperparameters, and training logs).

\subsubsection{Reproducing Bugs from Github Issues}
To reproduce the bugs from GitHub issues, we employed a systematic approach. First, we located the bug-inducing commit of a bug using the commit ID provided by the benchmark. Next, we cloned the corresponding repository and checked out the buggy version of the code using the commit ID. We then set up the development environment according to the reported bug, which may include installing the required version of the programming language and necessary dependencies or libraries, configuring environment variables and setting up any necessary databases, services, or external dependencies the code requires. If applicable, we applied necessary edit actions to the code, such as modifying specific lines or adding/removing code snippets, to trigger the buggy behaviour. We then ran the updated code snippet or the specific part of the codebase where the bug was expected to occur. To verify the presence of a bug, we compared the observed behaviour with the reported buggy behaviour and ran the test cases from the existing benchmark (Defects4ML). Finally, if the bug was reproduced, we recorded the edit actions used and critical information necessary for the bug reproduction.

\subsubsection{Agreement Analysis}
The first author and one independent collaborator conducted the bug reproduction process. We first reproduced 10 bugs from our dataset and achieved a Cohen Kappa of 54.5\%. Then, we had two meetings to identify the main reasons for our disagreements and resolved them. In the next round, we reproduced 10 more bugs and achieved a Cohen Kappa of 89.1\%, which is considered an almost perfect agreement~\cite{mchugh2012interrater}. After achieving an almost perfect agreement, the first author reproduced the remaining bugs (i.e., 128) while the independent collaborator checked the reproduction, achieving an average Cohen Kappa of 85.4\%. If the first author fails to reproduce the bug within 60 minutes, the independent collaborator also attempts to reproduce the bug. If both of them failed to reproduce the bug, the bug was marked as irreproducible. We spent $\approx$280 person-hours on the manual bug reproduction process.

\subsection{\textit{Verification of Bug Reproduction}} 
To verify the successful reproduction of bugs, we employed different strategies depending on the nature of the bugs. 

\subsubsection{Explicit Bugs}
An explicit bug results in an error message or exception. To verify the reproduction of this bug, we adopted a straightforward approach. We extracted the error message from the bug report and attempted to reproduce the bug under the reported conditions. We considered the bug reproduction successful when the observed error message matched the error message in the bug report.

\subsubsection{Silent Bugs}
Silent bugs are also referred to as functional or numerical errors. They do not result in system crashes or hangs and do not display error messages, but they lead to incorrect behaviour~\cite{tambon2024silent}. We adopted a comprehensive approach to verify the reproduction of silent bugs, as they often manifest subtly through silent issues like slow training or low accuracy.

We reproduced the silent bugs from Stack Overflow posts and GitHub issues. Subsequently, we used the evaluation metrics reported in the original posts or issues as the ground truth to verify the buggy behaviour. In cases where the code snippet was incomplete, we applied our edit actions to make it compilable, executable, and runnable.

To determine if a bug was successfully reproduced, we followed these steps:
\begin{itemize}
\item We executed the modified code snippet five times, each time with a different random seed, and calculated the average evaluation metric across these runs.
\item We compared the average evaluation metric to the reported evaluation metric of the buggy model.
\item If the average evaluation metric was within a 5\% error margin of the reported metric, we considered the bug to be reproduced.
\end{itemize}

We selected the 5\% threshold for error margin based on the existing literature~\cite{pham2020problems, alahmari2020challenges}. Pham et al.\cite{pham2020problems} found that implementation-level non-determinism could account for $\approx$3\% variance in the training and evaluation metrics of DL models, often caused by factors such as parallel processing issues, automatic selection of primitive operations, task scheduling, and floating-point precision differences. Furthermore, Alahmari et al.\cite{alahmari2020challenges} demonstrated that the variance in evaluation metrics could vary from 3\% to 7\% for models trained using the same dataset and code. Thus, our 5\% threshold accounts for the inherent variability in deep-learning models while still maintaining a reasonable standard for bug reproduction. This systematic approach allows us to verify the successful reproduction of silent bugs, ensuring the reliability of our findings. 

\subsubsection{Information Collection During Verification}
Following the successful reproduction of any bug above, we capture various information related to each bug, such as the deep learning architecture involved, edit actions used to reproduce the bug, time taken to reproduce the bug, type of bug reproduced, and the type of information present in the bug report. All these data gathered during bug reproduction helped us identify key edit actions and information components to reproduce specific types of deep learning bugs (Step 4, Fig.~\ref{fig:schematicDiagram}).

\subsection{\textit{Identifying Type Specific Information and Edit Actions}}

\subsubsection{Algorithm Selection}
To establish a relationship among the bug types, component information, and the key editing actions necessary for bug reproduction, we use the Apriori~\cite{apriori} algorithm. Apriori is a well-known algorithm for mining frequent itemsets from a list of transactions. It helps one identify common patterns and associations between different elements.

The Apriori algorithm exploits the principle that if an itemset is frequent across the transactions, then all of its subsets must also be frequent. It starts by identifying frequent individual items in the dataset and extends them to larger and larger itemsets as long as they meet certain constraints (e.g., support threshold). The algorithm terminates when no further successful extensions are found. More specifically, the Apriori algorithm consists of two main steps: the join step and the prune step. In the join step, the algorithm generates new candidate itemsets by joining the frequent itemsets found in the previous iteration. In the prune step, the algorithm checks the support count of each candidate itemset and discards the itemsets that do not meet the minimum support threshold. This process is repeated until no more frequent itemsets are generated. In our context, we apply the Apriori algorithm to analyze the information gathered during bug reproduction. Our goal was to determine the frequent combinations of bug types and component information, as well as edit actions during bug reproduction. We leveraged the Apriori algorithm's systematic pattern-mining capabilities to establish these relationships. By identifying the frequent itemsets, we can gain insights into the common patterns and associations among bug types, component information, and edit actions, which can help us understand and improve the bug reproduction process.

\subsubsection{Generating the Transactions}
To create our datasets for the Apriori algorithm, we employed a character encoding, where we encoded all labels into a unique character (e.g., `Training Bug' was encoded as `T', `Model Bug' was encoded as `M', `Obsolete Parameter Removal' was encoded as `O') and converted the data into transactions using the following format.

$$
\text{\textbf{Bug Type}} \rightarrow \text{\textbf{Information Category}}
$$
\textit{Example:} $T \rightarrow DH$ \\
\textit{Description:} The transaction above indicates that the reproduced bug is a training bug (T). The corresponding bug report contains useful information about the bug, such as the dataset used for training (D) and hyperparameters used by the model (H).

$$
\text{\textbf{Bug Type}} \rightarrow \text{\textbf{Edit Action}}
$$
\textit{Example:} $M \rightarrow OLN$ \\
\textit{Description:} The transaction above indicates that the reproduced bug is a model bug (M). To reproduce the bug, we performed three edit actions, as described below:
\begin{itemize}
    \item Obsolete Parameter Removal (O): We removed some of the parameters that were absent in the recent library and framework version to ensure that the code compiles.
    \item Logging (L): We logged various intermediate program states to verify the bug's presence.
    \item Neural Network Definition (N): We reconstructed the neural network based on the information provided in the bug report.
\end{itemize}

Following the specified format, we created two datasets of transactions that associate bug types with crucial information and edit actions, respectively. After creating the transactions, we use the Apriori algorithm to compute the support and confidence for our generated itemsets and association rules. We talk about these metrics in detail below.

\subsubsection{Metrics for Apriori Algorithm}
\textit{Support} is the proportion of transactions in the dataset that contain a particular itemset. Mathematically, the support of an itemset X is defined as the ratio of the number of transactions containing X to the total number of transactions. It is expressed as:

\[ \text{Support}(X) = \frac{\text{Transactions containing } X}{\text{Total number of transactions}} \]

For a rule \(X \Rightarrow Y\), where \(X\) and \(Y\) are two itemsets, the support is calculated for the combined itemset \(X \cup Y\).

\textit{Confidence} measures the likelihood that a rule \(X \Rightarrow Y\) holds. It is defined as the ratio of the support of the combined itemset \(X \cup Y\) to the support of the antecedent itemset \(X\). Mathematically, confidence is expressed as:

\[
\text{Confidence}(X \Rightarrow Y) = \frac{\text{Support}(X \cup Y)}{\text{Support}(X)}
\]

Confidence values range from 0 to 1. A high confidence value indicates a strong association between antecedents and consequent itemsets.

\subsubsection{Definitions for Apriori Algorithm}
\begin{itemize}
\item \textbf{Itemset:} An itemset is a set of one or more items. In our context, an item can be a bug type, an information category, or an edit action. For example, \{T, D, H\} is an itemset containing three items: bug type T (training bug), information category D (dataset), and information category H (hyperparameters).

\item \textbf{Transaction:} A transaction is a record that contains one or more items. In our study, we have two types of transactions as follows:
\begin{itemize}
\item Bug Type $\rightarrow$ Information Category: These transactions associate a bug type with the useful component information for reproducing that bug.
\item Bug Type $\rightarrow$ Edit Action: These transactions associate a bug type with the edit actions performed to reproduce that bug.
\end{itemize}

\item \textbf{Rule:} A rule is an implication of the form $X \Rightarrow Y$, where $X$ and $Y$ are itemsets. It suggests that if itemset $X$ is present in a transaction, then itemset $Y$ is likely to be present as well. In our study, rules are generated from the transactions to establish associations between bug types and component information or edit actions.
\end{itemize}

\subsubsection{Association Rule Generation}
We conducted two separate association rule mining operations in our study - one focused on component information while the other focused on edit actions used to reproduce deep learning bugs.  The Apriori algorithm generates rules by first identifying frequent itemsets and then creating rules from them. The steps below explain the process of rule generation with an example.
\begin{itemize}
\item \textbf{Identify frequent itemsets:} The algorithm scans the transactions to find itemsets that occur frequently while satisfying the minimum support threshold. For example, if the item {T, D} appears in 20\% of the transactions and the minimum support threshold is 10\%, it is considered a frequent item.

\item \textbf{Generate rules:} Once frequent itemsets are identified, the algorithm generates rules from them. For each frequent itemset, the algorithm creates rules by splitting the itemset into antecedent (left-hand side) and consequent (right-hand side). For example, from the item {T, D, H}, the following rules can be generated:
\begin{itemize}
\item $T \Rightarrow D, H$
\item $D \Rightarrow T, H$
\item $H \Rightarrow T, D$
\item $T, D \Rightarrow H$
\item $T, H \Rightarrow D$
\item $D, H \Rightarrow T$
\end{itemize}
\item \textbf{Calculate confidence:} For each generated rule, the algorithm calculates the confidence value. Confidence measures the likelihood that the consequent itemset appears in a transaction given that the antecedent itemset is present. Rules with confidence values above a minimum threshold are considered strong associations.
\end{itemize}
In particular, we extracted 27 itemsets and 34 rules for component information, highlighting the useful information for reproducing deep learning bugs. Similarly, we extracted 126 itemsets and 284 rules for edit actions, capturing the edit actions needed to reproduce bugs. We generated association rules based on the entries in our dataset and did not filter or remove any rules before determining confidence values. We then calculate the confidence values for all generated rules to identify the most influential ones for connecting bug types with edit actions and useful information. 

\subsubsection{Computation of Confidence Values for the Generated Association Rules}
As discussed earlier, support indicates how frequently a rule occurs, while confidence indicates the generality of the rule. To compute the confidence values for each association rule, we performed the following steps:
\begin{itemize}
\item \textbf{Calculate Support for Antecedent (\(X\)):} We calculate the support for the antecedent, which is the bug type in our case (e.g., `T' for Training Bug or `M' for Model Bug). Support for \(X\) is the proportion of all transactions that contain the specific bug type.
\item \textbf{Calculate Support for Combined Itemset (\(X \cup Y\)):} We then calculate the support for the combined itemset, \(X \cup Y\), which includes both the bug type and the information category or edit action (e.g., `T \(\cup\) H' for Training Bug associated with Hyperparameter Information, or `M \(\cup\) O' for Model Bug associated with Obsolete Parameter Removal).
\item \textbf{Compute Confidence for the Rule (\(X \Rightarrow Y\)):} The confidence of the rule \(X \Rightarrow Y\) is computed by dividing the support of the combined itemset \(X \cup Y\) by the support of the antecedent \(X\). This step gives us the confidence value, which indicates how often the information category or edit action \(Y\) is associated with the bug type \(X\) in our transactions.
\end{itemize}
For example, consider the rule \(T \Rightarrow D\). This rule indicates that if the type of bug is a `Training Bug' (T), it can be reproduced by the edit action `Input Data Generation' (D). To calculate the confidence of this rule, we count the number of transactions in which the training bug is reproduced by using the edit action `Input Data Generation' and divide this count by the total number of transactions involving training bugs in the dataset.

\subsubsection{Identification of High Confidence Associations}
We use high confidence and support values to detect the rules that reliably capture the core factors necessary to reproduce specific types of deep learning bugs. Based on these high-confidence rules, we identify the top 3 pieces of useful information and the top 5 edit actions used to reproduce each bug type. The decision to select three useful pieces of information and five edit actions was influenced by two key factors. First, we adhere to the Parsimony Principle~\cite{gori2023machine}, which suggests that selecting the simplest set of rules is preferable when multiple rules can predict or describe the same phenomenon. We thus concentrate on the most significant factors by selecting the top 3 and top 5 rules for edit actions and useful information, respectively. Second, we filter the rules based on minimum confidence values of 30\%, as suggested by Liu et al.~\cite{liu1999mining}. With our limited dataset, a 30\% confidence threshold can indicate a substantial pattern since finding associations in 30\% of cases points to a meaningful correlation given the data size. Furthermore, the 30\% minimum confidence helps filter out spurious correlations with small datasets that can occur by chance. Therefore, for our dataset, the selected threshold strikes an effective balance - it is high enough to identify meaningful associations in the data while eliminating noise from false correlations. Overall, this filtration left us with 23 rules for edit actions and 20 for useful information. Focusing on these high-confidence, high-support rules can reveal the patterns that reproduce deep learning bugs (Step 5a, 5b, Fig.~\ref{fig:schematicDiagram}).

\subsection{\textit{User Study}} To assess the benefits and implications of our findings in a real-life setting, we conduct a user study involving 22 developers (10 from academia + 12 from industry) (Step 6, Fig.~\ref{fig:schematicDiagram}). We discuss our study setup, including instrument design and participation selection, as follows:

\looseness=-1
\textbf{Instrument Design:} We used Opinio, an online survey tool recommended by our institution, to construct and distribute our questionnaire. Opinio enabled us to track the time spent by the participants on each individual question, which proved to be useful for our further analysis. The use of Opinio also did not require any additional effort from the participants, which made it a suitable choice for our user study. We divided our questionnaire into three sections. We discuss them in detail below:
\begin{itemize}
	\item \textbf{Introduction}: We first summarize our findings on the reproducibility of deep learning bugs to provide the participants with the necessary context and background information. For the survey itself, we do not give the respondents a fixed time to complete it, but we specify that it should take $\approx$60 minutes on average; this number was derived from our pilot study. This was done to ensure that the respondents do not work under time pressure. Since the participants have different levels of experience, allocating a fixed time for bug reproduction might affect our results.
	\item \textbf{Demographic Information}: After providing the contextual information about our study, we collect demographic information from developers (e.g., experience bug fixing in deep learning frameworks). We then ask the developers to elaborate on the challenges that they face when reproducing deep learning bugs in their daily lives.
	\item \textbf{Questionnaire Preparation}: First, we select eight bugs (2 Tensor, 2 API, 2 Model, and 2 Training Bugs) from our dataset constructed during manual bug reproduction and dataset creation. During this process, we categorized the bugs by difficulty level and type based on the number of edit actions and critical information required to reproduce them. We use stratified random sampling to pick 2 bugs from each type; one of the bug types is relatively easy to reproduce with only 1 edit action, and the other type is relatively difficult to reproduce warranting multiple edit actions. We pick 8 bugs following this approach and, then randomly assign them to four sets, each containing 1 easy and 1 difficult bug. Second, we provide the users with the issue description and the code snippet from the original Stack Overflow post. We also provide a Google Colaboratory notebook containing the code for sample edit operations to aid the bug reproduction process. Finally, we include a free-text box in the form to allow users to share any additional information about edit actions not covered in our study.
\end{itemize}  

\textbf{Study Session:} During our study session, each participant completes the following five tasks. First, the participant provides their demographic information. Second, the participant explains their daily challenges when reproducing deep learning bugs. Third, each participant reproduces two deep learning bugs and self-reports the edit actions and information they used to reproduce the bugs. Fourth, the participant also provides the rationale behind their self-reported edit actions and component information used to reproduce the bugs. Finally, the participant provides information about any other edit actions that they might have used to reproduce the bugs but are not covered in our study.

\textbf{Participant Selection:} We first conducted a pilot study with two researchers and two developers. Based on their feedback, we rephrased ambiguous questions and added sample code to aid the manual bug reproduction by the users. Incorporating this constructive feedback enabled us to refine and improve the quality of our final questionnaire. Then, we invite professional developers and researchers with relevant deep learning experience to our study. We send our invitations to the potential participants using direct correspondences, organization mailing lists (e.g., Mozilla Firefox), and public forums (e.g., LinkedIn and Twitter). A total of 22 participants responded to our invitations. Out of them, 10 (45\%) came from academia, and 12 (55\%) came from industry. In terms of bug-fixing experience with deep learning bugs, 14 participants (63.63\%) had 1-5 years of experience, 4 (18.18\%) had 5-10 years of experience, and the remaining 4 (18.18\%) had less than one year of experience. In terms of deep learning frameworks, 19 participants (86.34\%) reported having working experience with Tensorflow, 21 (95.45\%) reported experience with PyTorch, and 20 (90.91\%) reported experience with Keras. All these statistics indicate a high level of cross-framework expertise within our participants. 

\textbf{Defining Control and Experimental Groups:} We carefully divided the study participants into control and experiment groups, as shown in Table.~\ref{tab:developerDistributionUserStudy}. Using these two groups, we wanted to assess the benefit of our recommended information in the context of bug reproduction. The control group receives no hints about how to reproduce a bug. In contrast, the experimental group receives hints (e.g., useful information, edit actions) that could help them reproduce a bug. 

\textbf{Ensuring Similar Experience Levels:} Our guiding principle was to ensure that both groups had a similar distribution in their relevant experience. Specifically, we surveyed all the developers about their experience and used a stratified random sampling approach to assign them to the two groups. This randomization allowed us to minimize potential bias and confounding factors across the groups.

\begin{table}[h]
	\centering
	\caption{Distribution of participants in the control and experimental groups}
	\resizebox{\columnwidth}{!}{%
		\begin{tabular}{|l|l|l|}
			\hline
			\textbf{DL Experience} & \textbf{Developers in Control Group} & \textbf{Developers in Experimental Group} \\ \hline
			\hline
			\textbf{\textless{}1 Year} & 3 (27.27\%) & 3 (27.27\%) \\ \hline
			\textbf{1-5 Years}         & 6 (54.54\%) & 5 (45.45\%) \\ \hline
			\textbf{5-10 Years}        & 2 (18.18\%) & 3 (27.27\%) \\ \hline
		\end{tabular}%
	}
	\label{tab:developerDistributionUserStudy}
\end{table}

\textbf{Leveraging Associations to Produce Hints for User Study:} We use high-confidence associations from our RQ2 (Step 3.6.6) to develop hints for our user study. These associations revealed the edit actions and critical information needed to reproduce specific types of bugs. We use the steps below to construct the hints in a systematic way for our user study.

\begin{enumerate}
\item \textbf{Bug Categorization:}
\begin{enumerate}
\item Analyze the information found in a bug report to identify the specific type or category of the bug.
\item Use the identified bug category to retrieve relevant component information and edit actions.
\end{enumerate}

\item \textbf{Retrieve component information and edit actions:}
\begin{enumerate}
\item Based on the identified bug category, retrieve the top 3 components and top 5 edit actions from the findings of RQ2.
\item The component information includes the most important aspects required to reproduce or understand a bug (e.g., shape of input data, training code, error messages).
\item The edit actions refer to the most frequently associated actions to a specific bug category.
\end{enumerate}

\item \textbf{Determining the Most Relevant Statement:}
We also collect the most relevant statement from the bug description as follows.
\begin{enumerate}
\item Split the bug report's text into a list of sentences using '.' as the delimiter.
\item Collect a list of key phrases for the components retrieved in Step 2(a). These key phrases are derived from our qualitative analysis of the bug reports and can be found in the replication package~\cite{replicationPackage}.
\item Generate the Sentence-BERT embeddings for every statement in the bug report and the keywords above using the 'sentence-transformers\-/all-MiniLM-L6-v2' pre-trained model.
\item For each statement in the bug report, calculate the cosine similarity between the statement embedding and the embeddings of the keywords.
\item Identify the relevant statement for the most prevalent component (i.e., top component from Step 2(a)) based on their cosine similarity score.
\item Prepare a context-specific hint that guides the user to a particular statement using the following template: `Focus on the statement: $<$Statement extracted in Step 3(e)$>$'
\end{enumerate}

\item \textbf{Hint Formulation:}
\begin{enumerate}
\item Combine the retrieved component information and edit actions from the research findings with the context-specific hint to formulate the complete hint for the user study.
\item Use the template below to formulate the hint for each bug.
\end{enumerate}
\end{enumerate}
\begin{tcolorbox}[title=Template for Hint Generation, label=hintbox]
\textbf{Hints} \\
1. \texttt{$<$CI1$>$}, \texttt{$<$CI2$>$}, \texttt{$<$CI3$>$} can be useful information for reproducing the bug. \\
2. \texttt{$<$EA1$>$}, \texttt{$<$EA2$>$}, \texttt{$<$EA3$>$}, \texttt{$<$EA4$>$}, \texttt{$<$EA5$>$} can be useful edit actions for reproducing the bug. \\
3. Focus on the Statement: \textbf{``$<$Most Relevant Statement from the SO Post$>$"}
\end{tcolorbox}
We ensure a systematic formulation of hints for our study by following the steps above. This approach incorporates high-confidence associations from Step 3.6.6 and the bug description for individual bug. With this intervention, we plan to measure if and how our recommended information help participants reproduce the bugs more accurately or quickly. The code and results for the hint formulation are available in our replication package~\cite{replicationPackage}.

\subsection{\textit{User Study Results Analysis}}

To assess the effectiveness of our edit actions, we analyze the participants' responses and qualitative feedback. Specifically, we analyze the edit actions used by the participants and compare them with those used in our manual bug reproduction process. If similar edit actions were used, it would indicate that our recommended edit actions and component information were effective in the reproduction of the bugs. 

In our user study, the participants first reproduce their assigned bugs and report their used information or actions to reproduce their bugs. To analyze the effectiveness of our recommended information, we compare the control and experimental groups in terms of their success rates in bug reproduction and the time taken to reproduce the bugs. We select bug reproducibility rate and time taken for bug reproduction as our key metrics to evaluate the effectiveness of our findings. These quantitative metrics directly measure how successful and efficient our recommended information is in assisting the developers to reproduce deep learning bugs. Higher reproducibility rates and reduced reproduction times in the experimental group compared to the control group would indicate that our findings are effective for improving deep learning bug reproducibility. We also analyze the qualitative feedback from the participants to determine if our provided hints were useful or not. We also analyze them to uncover new insights into deep learning bug and their reproduction.

By analyzing the user study results and the developers' qualitative feedback, we gain new and valuable insights into the reproducibility of deep learning bugs. The user study results and feedback also provide information about the effectiveness of our recommended actions for bug reproduction. This enables us to deliver novel and actionable insights regarding the current state of deep learning bug reproducibility based on empirical evidence (Step 7, Fig. \ref{fig:schematicDiagram}).

\begin{table}
\captionsetup{justification=raggedright,singlelinecheck=false} 
\caption{Summary of the Reproduced Bugs}
\label{tab:reproductionStatistics}
\resizebox{\columnwidth}{!}{%
\begin{tabular}{|l|l|l|}
\hline
{\color[HTML]{000000} \textbf{Type of Bug}} & {\color[HTML]{333333} \textbf{Bugs Reproduced}} & {\color[HTML]{000000} \textbf{Model Architectures Covered}} \\
\hline
Training (T)        & 50 & CNN, LSTM, AutoEncoder, MLP, RCNN, \\ & & ResNet           \\ \hline
Model (M)           & 42 & BERT, CNN, GMM, LSTM, MLP \\
& &  VGG16, Transformers      \\ \hline
API (A)             & 20 & CNN, GAN, MLP, Transformers, VGG19, \\ 
& & Variational RNN \\ \hline
GPU (G)             & 3  & -                                                   \\ \hline
Tensor and Input (I) & 29 & CNN, GAN, Logistic Regression, MLP, ResNet          \\ \hline
Mixed (X)           & 4  & CNN, BERT, MLP \\ \hline
\end{tabular}%
}
\end{table}
\section{Study Findings} \label{section5}
In this section, we present the findings of our study by answering three research questions as follows.
\subsection{\textit{RQ1: Which edit actions are crucial for reproducing deep learning bugs?}}
To answer the first research question, we worked with 165 bugs, and reproduced 148 bugs of different types and architectures. Table.~\ref{tab:reproductionStatistics} briefly summarizes our reproduced bugs. Through our comprehensive reproduction process, we identify ten edit actions that are crucial for reproducing deep learning bugs. Table.~\ref{tab:editActions} shows the identified actions from our qualitative analysis. We explain these actions as follows.

\textbf{Input Data Generation (A$_{1}$)} is one of the key edit actions for reproducing deep learning bugs. This action involves programmatically generating synthetic input data that closely matches the characteristics of the original data used for training the model. The key objective of input data generation is to simulate representative data that can trigger or reproduce the erroneous model behavior described in the bug report. This allows for the reproduction of issues that manifest only in the presence of specific data properties or distributions. To perform input data generation, we leverage any details about the data that are provided in the bug report, such as data types, value ranges, shapes, distributions, preprocessing steps, etc. For example, for image data, the report may specify that inputs are RGB images of size 224x224x3 with pixel values normalized to [0,1]. Similarly, for text data, the description may indicate sequences of 512 tokens processed using a particular tokenizer. Using this information, we can systematically generate synthetic data matching the properties through appropriate library functions. For images, we can use libraries like OpenCV~\cite{OpenCV2024} or PIL~\cite{PIL} to construct random images of the required size and channels. For text, we can sample token sequences from a standard corpus or use specialized generative models like GPT-2~\cite{radford2019language}. Our manual bug reproduction shows that $\approx$73\% of our collected posts from Stack Overflow have the relevant data characteristics. Let us consider the issue reported in the Stack Overflow post (Issue \#61781193). In this post, the reporter suspects that the model is not learning - as evidenced by the constant training loss across the epochs. The following text from the post shows how the reporter might submit the input data distribution. 

\textit{R: My training data has input as a sequence of 80 numbers in which each represent a word and target value is just a number between 1 and 3.}

Using this information, we generated the random input data as follows and were able to reproduce the corresponding bug.

\begin{lstlisting}
train_data = torch.utils.data.TensorDataset(torch.randint(0, 200, (1000, 80)), torch.randint(1, 3, (1000,)))
\end{lstlisting}

\textbf{Neural Network Construction (A$_{2}$)} was one of the most used edit actions during our bug reproduction. In this edit action, we construct a neural network based on the architecture provided by the reporter. Similar to the data characteristics, the information about the neural network is present in $\approx$65\% of the reproducible issue reports. Using the neural network description from the issue reports, we were able to construct the models. Let us consider the issue reported in the Stack Overflow post (Issue \#63204176). In this post, the reporter submits an issue where a CrossEntropy loss function within a loop is overwritten with a Tensor, causing a TypeError in later iterations. When we analyze the post, we find that the reporter mentions that they have used a logistic regression model (1-layer neural network with a sigmoid activation function~\cite{logisticRegression}). However, the reporter does not provide the code snippet necessary for reproducing the bug.

\textit{R: I am trying to write a simple multinomial logistic regression using mnist data.}

To reproduce the bug, we constructed multiple logistic regression models for the MNIST dataset, as highlighted in the post. We added the import statements, wrote the code to load the MNIST dataset, and completed the code. Using the training loop's code snippet from the original code and our edit actions, we could complete the code snippet and reproduce the bug successfully. Using this partial information, we constructed multiple multinomial logistic regression models, and despite the lack of relevant code examples, we successfully reproduced the corresponding bug.
\begin{table*}
\captionsetup{justification=raggedright,singlelinecheck=false} 
    \caption{Edit Actions for Reproducing Deep Learning Bugs}
    \begin{tabularx}{\textwidth}{p{0.5\textwidth}|X}
        \hline
        \textbf{Edit Action} & \textbf{Overview} \\
        \hline
        \hline
        \textit{A$_{1}$}: \textit{Input Data Generation (D)} & Generating input data that simulates the data used for training the model. \\ \hline
        \textit{A$_{2}$}: \textit{Neural Network Construction (N)} & Reconstructing or modifying the neural network based on the information provided. \\ \hline
        \textit{A$_{3}$}: \textit{Hyperparameter Initialization (H)} & Initializing the hyperparameters for training, such as batch size and number of epochs. \\ \hline
        \textit{A$_{4}$}: \textit{Import Addition and Dependency Resolution (R)} & Determining the dependencies in the code snippet and adding the missing import statements. \\ \hline
        \textit{A$_{5}$}: \textit{Logging (L)} & Adding appropriate logging statements to capture relevant information during reproduction. \\ \hline
        \textit{A$_{6}$}: \textit{Obsolete Parameter Removal (O)} & Removing outdated parameters or functions to match the parameters of the latest library versions. \\ \hline
        \textit{A$_{7}$}: \textit{Compiler Error Resolution (C)} & Debugging and resolving compiler errors that arise due to syntactic errors in the provided code snippet. \\ \hline
        \textit{A$_{8}$}: \textit{Dataset Procurement (P)} & Acquiring the necessary datasets and using them to train the model.  \\ \hline
        \textit{A$_{9}$}: \textit{Downloading Models \& Tokenizers (M)} & Fetching pre-trained models and tokenizers from external sources. \\ \hline
        \textit{A$_{10}$}: \textit{Version Migration (V)} & Updating code to adapt to changes introduced in newer versions of libraries and frameworks. \\ 
        \hline
    \end{tabularx}
    \label{tab:editActions}
\end{table*}

\textbf{Hyperparameter Initialization (A$_{3}$)} is one of the core edit actions for reproducing deep learning bugs. As a part of this edit action, we initialize various hyperparameters (e.g., number of epochs, batch size, optimizer) for training the neural networks. Sometimes, the reporter did not provide these configurations, and we had to initialize them using default parameters to reproduce the bug. This is further evidenced by the fact that only $\approx$53\% of our collected posts from Stack Overflow have information about the hyperparameters. Let us consider the issue reported in the Stack Overflow post (Issue \#31880720), where the author gets a poor test accuracy of 13.9\% when training a neural network on a synthetic binary classification dataset. After manual analysis of the post and the code snippet, we observe that some of the hyperparameters in the code snippet have not been initialized.

\begin{lstlisting}
model.fit(X_train, Y_train, batch_size=batch_size, nb_epoch=nb_epoch, show_accuracy=True, verbose=2, validation_data=(X_test, Y_test))
\end{lstlisting}

To reproduce the bug, we initialized the batch size with commonly-used values $\{32, 64, 128\}$ and the number of epochs ranging from 1 to 10. Since we initialized these hyperparameters, we ran the edited code snippet five times to confirm the effectiveness of our edit operation. We were able to reproduce the bug in all five iterations. We performed the steps mentioned above for all 56 bugs involving hyperparameter initialization.

\textbf{Import Addition and Dependency Resolution (A$_{4}$)} are one of the most common edit actions for all types of bugs. In this edit action, we analyze the code manually and determine the dependencies required for the code snippet to run. Then, we install the dependencies and manually import them to complete the code snippet. This edit action was also used to reproduce traditional software bugs, as reported by Mondal et al.~\cite{msr2019mondal}. In the Stack Overflow posts we collected, $\approx$47\% of the issue reports lacked the required dependency information. This significant absence of the dependency information further highlights the need for our suggested edit action. Let us consider the issue reported in the Stack Overflow post (Issue \#50306988). In this post, the neural network model with softmax activation struggles to fit a simple 2-feature classification dataset, converging extremely slowly compared to logistic regression, which achieves 100\% accuracy. However, the code snippet reported in the post was incomplete, and the dependency details were missing. Hence, we resolved the dependencies (keras, numpy and random) and added the required import statements to the code snippet. With the resolved dependencies, we were able to reproduce the bug successfully.

\textbf{Logging (A$_{5})$} plays a crucial role in the reproduction of different types of bugs. This action involves adding log statements to the provided code snippet to verify the reproduction of a bug. Let us consider the issue reported in the Stack Overflow question (Issue \#70546468). The reporter provides a numpy array with shape (1, 3). The reporter then vertically concatenates (stack depthwise) multiple copies of this array to create tensor groups with shapes (2, 3) and (1, 3). After this vertical concatenation process, the reporter expects the tensor's shape to be (2, None, 3). Unfortunately, the tensor's shape was (2, None, None), according to the reporter. To verify the claim made by the reporter, we introduce a log statement within the code snippet. When we run the modified code snippet, we observe the shape of the stacked tensor to be (2, None, None), thereby confirming the reproduction and validity of the bug.

\textbf{Obsolete Parameter Removal (A$_{6}$)} is a crucial edit action that enhances bug reproducibility when differences in library and framework versions cause compatibility issues. It involves removing obsolete parameters no longer supported by newer versions. Frequent updates to deep learning frameworks and APIs can cause breaking changes, i.e., the code written for older versions of the framework is incompatible with newer versions. Developers reproducing these bugs often face challenges when the bug report's environment is significantly older than their current working environment, and downgrading to match those outdated versions is not always feasible due to various constraints. In some cases, the bug report uses significantly outdated versions that are no longer supported or maintained by the framework developers. Additionally, the developer's current project may have dependencies that require newer versions of Python or the frameworks, making it impractical to downgrade. Furthermore, organizations or development teams may have policies in place that mandate using the latest stable versions for security, performance, and maintainability reasons, preventing the use of older versions.

In such cases, developers must port the bug-reproducing code to their current environment, where Obsolete Parameter Removal edit action proves highly beneficial. It makes the code compatible with newer framework versions while preserving the bug-reproducing behaviour. Consider the issue reported in the Stack Overflow post (Issue \#65992364). In this post, the reporter attempts to optimize an object detection model using `pytorch-mobile', but the code snippet fails to optimize the model file size. The code snippet in the issue also contains the following line of code. 

\begin{lstlisting}
script_model_vulkan = optimize_for_mobile(script_model, backend=``Vulkan")
\end{lstlisting}
However, according to the API documentation of PyTorch 1.6.0~\cite{pytorch1.6}, the \verb|backend| parameter was no longer supported. We removed the obsolete parameter and thus were able to reproduce the bug.

\textbf{Compiler Error Resolution (A$_{7}$)} is one of the extensively employed edit actions in reproducing bugs. In this edit action, we resolve compiler errors to get the code running and reproduce the bug. While downgrading the compiler or library versions can sometimes resolve errors or enable the bug reproduction with deprecated functionality, it may not always be feasible. For example, major framework releases often remove support for older versions, making downgrading impossible. Additionally, Python version mismatches between the original buggy code and the current development environment can prevent successful downgrading. In cases where downgrading is infeasible due to such constraints, correction of compiler errors enables reproducing deep learning bugs despite incompatible environments.

For example, in the Stack Overflow post (Issue \#71514447), the reporter states that the training loss is significantly increasing after every epoch. They suspect that the bug might be due to the incorrect computation of loss values. Furthermore, to help the developers reproduce the bug, they provide a detailed description of the bug and a complementary code snippet. However, the code snippet provided by the user does not compile, as the \verb|criterion| function has not been defined. Hence, we replaced the function parameter with the default value of cross-entropy loss, as shown below.
\begin{lstlisting}
def train_model(model, optimizer, train_loader,  num_epochs, criterion = nn.CrossEntropyLoss()):
\end{lstlisting}
Using the edit action above, we resolved the compiler errors and thus were able to reproduce the corresponding bug.

\textbf{Dataset Procurement (A$_{8}$)} is a critical edit action to reproduce bugs that require specific datasets. In this edit action, we analyze the issue report and attempt to procure the dataset mentioned in the report. For instance, in the Stack Overflow post (Issue \#73966797), the reporter mentioned that they used the CIFAR-10, a well-known dataset for object detection. To reproduce the bug, we downloaded the dataset from its original source\footnote{https://www.cs.toronto.edu/~kriz/cifar.html}, and using the dataset and the training code, we could reproduce the bug.

\textbf{Downloading Models \& Tokenizers (A$_{9}$)} is one of the edit actions that was frequently used to reproduce bugs in Large Language Models and Transformer-based architectures. In this edit action, we download the pre-trained models and tokenizers for reproducing a bug. Let us consider the issue reported in the Stack Overflow post (Issue \#69660201). In this post, the reporter faces a ValueError when fitting a text classification model with a BERT tokenizer, due to a mismatch between the model's expected input and the tf.data.Dataset created from the text corpus and labels. For example, in the Stack Overflow post (Issue \#69660201), the reporter has provided the following information.

\textit{R: In the preprocessing layer, I'm using a BERT preprocessor from TF-Hub.}

Based on the above information, we added the relevant URLs and configured the code to download the preprocessor and encoder. After downloading the preprocessor and encoder, we successfully reproduced the corresponding bug.

\begin{lstlisting}
tfhub_handle_preprocess = ``https://tfhub.dev/tensorflow/bert_en_uncased_preprocess/3"
tfhub_handle_encoder = ``https://tfhub.dev/tensorflow/small_bert/bert_en_uncased_L-4_H-512_A-8/1"
\end{lstlisting}

\textbf{Version Migration (A$_{10}$)} is a vital edit action for reproducing bugs in deep learning frameworks and libraries. It involves adapting code written for older versions to the latest version, confirming the same bug's presence in the current environment. Mondal et al.~\cite{msr2019mondal} introduced a similar concept called `Code Migration'. Similar to Obsolete Parameter Removal, Version Migration aims to address compatibility issues caused by the updates in deep learning frameworks and APIs. While Obsolete Parameter Removal focuses on removing deprecated parameters, Version Migration contains additional modifications required to make the code follow the newer version's syntax, APIs, and functionalities. This may involve updating function calls, class constructors, import statements, and other code elements affected by the version changes. 

The necessity for Version Migration often arises when the versions used in the bug report are significantly outdated compared to the developer's current working environment. Bug reports may contain code snippets written for older framework versions that are no longer actively maintained or supported by the developers. In such cases, downgrading to the exact reported version is not feasible, as it would require reverting to unsupported and potentially insecure versions. Additionally, the developer's current project may have dependencies that require newer versions of Python or deep learning frameworks, making it impractical to downgrade. Furthermore, organizations or development teams may have policies in place that mandate using the latest stable versions for security, performance, and maintainability reasons, preventing the use of older versions. To demonstrate the utility of the Version Migration, let us consider the issue reported in the post on Stack Overflow (Issue \#45711636). The post describes an issue with the CNN architecture constructed by the user. When the user passes the input through the CNN, they encounter a ValueError, which highlights a problem with the negative dimension value of the input. However, the code snippet provided in the issue uses Tensorflow 1.3.0. To reproduce this issue in TensorFlow 2.14.0, we carefully migrated the model built in TensorFlow 1.3.0 to the syntax of Tensorflow version 2.14.0. Specifically, we updated the Sequential, Conv2D and MaxPooling2D constructors to match the TensorFlow 2.14.0 API syntax. We also changed padding and pool size parameter schemes and removed the obsolete input shapes. Finally, we upgraded the model compilation and imported the required Tensorflow 2.14.0 modules. After modifying the code snippet, we successfully reproduced the bug in our runtime environment. We verified the reproduction by comparing the error message from our updated code snippet with the one reported in the Stack Overflow post.

\begin{shaded}
\noindent \textbf{Summary of RQ1}: By manually reproducing \textbf{148} deep learning bugs, we identify \textbf{ten key edit actions} that could be useful to reproduce deep learning bugs (e.g., input data generation, neural network construction, hyperparameter initialization). These edit actions can help developers complete the code snippets and thus reproduce their deep learning bugs.
\end{shaded}
\subsection{\textit{RQ2: What component information and edit actions are useful for reproducing specific types of deep learning bugs?}}

\subsubsection{Identifying useful Information in Bug Reports} \label{usefulInformation}

While reproducing deep learning bugs, we kept track of information from bug reports that helped us reproduce them. After reproducing 148 bugs successfully, we have identified five useful pieces of information that can improve the chance of reproducing a bug. We discuss these factors in detail below.

\textbf{Data (F$_{1}$):} Data is one of the essential factors in ensuring the reproducibility of deep learning bugs. Deep learning systems heavily rely on the data~\cite{breck2019data}, and reproducing deep learning bugs becomes easier with access to the original data. Data helps us reproduce the deep learning bugs by providing the exact sample inputs that trigger the erroneous behaviour. By understanding the training data distributions and ranges, we can reconstruct the original training environment, which is crucial for reproducibility. However, the issue reports often lack direct information about the data. To address this problem, we collect information about the data by extracting the shape of the data, data distribution, type of variables and their corresponding ranges from the issue description. Leveraging this information and our proposed edit actions, namely Input Data Generation (A$_{1}$) and Dataset Procurement (A$_{8}$), we can generate or obtain the necessary data to reproduce deep learning bugs. According to our investigation, $\approx$77\% of our reproduced bugs contained information about the data (see Table.~\ref{tab:criticalInformationStatistics}) in their issue reports. For example, in the Stack Overflow post (Issue \#43464835), the reporter has provided the dimensions and sample row of the training dataset, as shown below.

\textit{R: I have a train dataset of the following shape: (300, 5, 720)}

\textit{Sample Input: [[[   6.   11.  389. ...,   0.    0.    0.]]]]
}

Using this information and our proposed edit action Input Data Generation (A$_{1}$), we first generated a data frame of size (300, 5, 720) that contains values in the range from 0 to 400. Then, by leveraging the generated data and other useful information, we successfully reproduced the corresponding bug.

\textbf{Model (F$_{2}$):} The model architecture describes the components of a deep learning model and how they transform inputs into outputs. Understanding the model architecture is crucial for reproducing deep learning bugs, as it provides insights into the model's components, connectivity, and the required neural network architecture. According to our investigation, $\approx$58\% of our reproduced bugs contained information about the model architecture (see Table.~\ref{tab:criticalInformationStatistics}) in their issue reports. However, the complete source code implementing the model architecture might not always be available in the issue reports. To overcome this limitation, we carefully gather information about a model's architecture, i.e., the number of layers, layer properties and activation function from the issue description. Leveraging this information and our proposed edit action Neural Network Construction (A$_{2}$), we reconstruct the model and reproduce the deep learning bugs. For example, let us consider the following text from a Stack Overflow post (Issue \#63204176) that mentions the use of multinomial logistic regression on the MNIST dataset, as shown below. Unfortunately, the reporter fails to provide the code snippet for the model.

\textit{R: I am trying to write a simple multinomial logistic regression using mnist data.}

Despite the absence of the code snippet, the issue description provides useful hints about the model architecture and dataset. Using them and our edit action -- Neural Network Construction (A$_{2}$), we successfully reproduced the bug.

\textbf{Hyperparameters (F$_{3}$):} Hyperparameters play a crucial role in controlling the learning process and model behaviour during training. They encompass parameters such as learning rate, batch size, number of epochs, optimizer, regularization techniques, and loss functions. The specific values chosen for these hyperparameters significantly impact a model's performance, training process, and bug manifestation. Thus, reporting the complete set of hyperparameters is essential to help reproduce deep learning bugs. According to our investigation, 48\% of our reproduced bugs contain information about the hyperparameters used in their issue reports. For the 52\% of bug reports that did not include hyperparameter information, we used the Hyperparameter Initialization edit action (A$_{2}$) to reproduce the bugs by initializing the hyperparameters with default values. In the example Stack Overflow post (Issue \#65993928), we can see how hyperparameters can play a crucial role in reproducing the bug.
\\
\begin{lstlisting}
loss2 = (2 * (log_sigma_infer - log_sigma_prior)).exp() \ +((mu_infer - mu_prior)/ log_sigma_prior.exp()) ** 2 \ - 2 * (log_sigma_infer - log_sigma_prior) - 1

loss2 = 0.5 * loss2.sum(dim = 1).mean()
\end{lstlisting}

Since the reporter was using a custom loss function, it was vital for them to share the value of constants and the formula used to calculate the loss. The reporter's custom loss function, with constants 2 and 0.5 and its formula, was crucial for reproducing a bug in model training due to incorrect loss calculation, emphasizing the need to provide a complete set of hyperparameters.

\textbf{Code Snippet (F$_{4}$):} Code snippets are critical for reproducing bugs, as highlighted by the fact that 98\% of professional developers consider them an essential component of bug reproducibility~\cite{soltani2020significance}. They include the data preprocessing, data splitting technique, code used for training the model, and implementation of the evaluation metrics. However, from our manual bug reproduction, we observe that even though code snippets are present in $\approx$82\% of the bug reports, only 9.41\% of them can be used verbatim for bug reproduction. To address this limitation, we use several of our proposed edit actions, such as Import Addition and Dependency Resolution (A$_{4}$), Logging (A$_{5}$), Obsolete Parameter Removal (A$_{6}$), Compiler Error Resolution (A$_{7}$) and Version Migration (A$_{10}$). These edit actions help us fix the errors in the code and make the code compileable and runnable. \\

To demonstrate the importance of submitting a complete code snippet, let us consider the Stack Overflow post with Issue \#76186890. In this issue, a high-quality code snippet helps us reproduce the bug related to a complicated architecture (T5) without significant changes. The code snippet uses the T5 model from HuggingFace Transformers for text-to-text translation. It preprocesses the input text data, configures the target IDs to predict a specific answer, calculates loss and perplexity metrics, and trains the model. Having a complete and runnable snippet was helpful in reproducing the bug. It provides sufficient details like data preparation, loss calculation, model training, and evaluation to improve the ease of reproducing deep learning bugs, which were essential for bug reproduction.

\textbf{Logs (F$_{5}$):} Logs provide a real-time record of the model's behaviour during training and inference. Traditional software logs consist of information such as event logs and stack traces, whereas deep learning systems consist of compiler error logs, training error logs, and evaluation logs~\cite{Chen_Jiang_2022}. Sharing these logs is crucial for reproducing deep learning bugs as they allow us to verify if we can reproduce the same erroneous behaviour reported in the original issue. From our manual bug reproduction, we discover that $\approx$88\% of our reproduced bugs contain the necessary logs in their issue reports. Such a high presence of logs in our dataset of reproduced bugs highlights the importance of logging in deep learning bug reproduction. By matching the logs from the original issue and the logs from reproduction on our local machines, we were able to confirm the presence of several bugs in the deep learning systems.

The Stack Overflow post (Issue \#34311586) shows how logs can be used to confirm the presence of deep learning bugs. In this particular issue, the reporter shared the training logs and the code snippet. We modified the code snippet with our proposed edit actions to make it compilable and runnable. When executed, we observed the same anomalous behaviour in the training logs as described by the original reporter. This demonstrates the importance of sharing training and evaluation logs in the issue report. We also found them to be one of the most useful pieces of information to reproduce deep learning bugs.

\begin{table}[]
\captionsetup{justification=raggedright,singlelinecheck=false}
    \caption{Prevalence of Useful Information in Reproducible Issue Reports}
    \label{tab:criticalInformationStatistics}
    \begin{tabular}{|c|c|c|c|c|c|c}
        \hline
        \textbf{Factor} & \textbf{Data} & \textbf{Model} & \textbf{Hyperparameters} & \textbf{Code Snippet} & \textbf{Logs} \\
        \hline
        \hline
        \textbf{Prevalence} & 77.4\% & 58.1\% & 47.9\% & 82.1\% & 87.6\% \\
        \hline
    \end{tabular}
\end{table}

\begin{table}[]
\centering
\caption{Top 3 Useful Component Information for Reproducing Specific Types of Deep Learning Bugs}
\label{tab:criticalInformationBugType}
\resizebox{0.65\columnwidth}{!}{%
\begin{tabular}{l|l}
\hline
\multicolumn{1}{l|}{\textbf{Training}} & \textbf{Model} \\ \hline \hline
\multicolumn{1}{l|}{Code Snippet (0.86)} & Logs (0.7857)           \\ \hline
\multicolumn{1}{l|}{Data (0.82)}         & Code Snippet (0.7143)          \\ \hline
\multicolumn{1}{l|}{Logs (0.76)}         & Model (0.6429)   \\ \hline
\textbf{Tensor}                        & \textbf{API}   \\ \hline \hline
Data (0.9655)                              & Logs (0.85)         \\ \hline
Logs (0.9310)                              & Code Snippet (0.75) \\ \hline
Code Snippet (0.7241)                      & Model (0.70) \\ \hline
\end{tabular}%
}
\end{table}

\subsubsection{Relationship between Useful Information and Type of Bugs}

During our manual analysis (Section~\ref{usefulInformation}), we identify useful information for reproducing deep learning bugs. We used the Apriori algorithm to determine the relationship between the type of bug and the information required to reproduce the bug. The insights from this analysis could serve two purposes -- detecting missing information in submitted bug reports and formulating follow-up questions to obtain any missing information needed for reproducibility. Table.~\ref{tab:criticalInformationBugType} summarizes our findings, and we discuss them in detail below.

\textbf{Data:} The accurate reproduction of bugs in deep learning systems heavily relies on the presence of data and its characteristics. They play a crucial role in reproducing two key categories of bugs - \textbf{training} bugs and \textbf{tensor} bugs. 

For training bugs, the data has a confidence value of 82.00\% according to our generated rules (check Table.~\ref{tab:criticalInformationBugType}). This high confidence highlights the significance of data in reproducing the numerous training issues that can manifest during model development. The details about the data, such as the number of samples, class distribution, feature distributions, data splitting ratios, and preprocessing steps, are instrumental in reproducing training bugs. For example, training a classification model on a highly skewed dataset is prone to overfitting. With the knowledge of the data distribution, we can recreate a representative dataset, which can help us reproduce the bugs, even if the actual dataset is not available. 

For tensor bugs, data characteristics have an even higher confidence of 96.55\%, according to our generated rules (Table.~\ref{tab:criticalInformationBugType}). This suggests that tensor attributes like shape, data type, sparsity, value ranges and origin are pivotal for reliably reproducing many bugs stemming from invalid dimensions or precision issues. For instance, bugs arising from tensor shape mismatches can emerge if the shape of the input data does not match the expected input shape. Furthermore, real-world data is often more vulnerable to human error and biases compared to synthetic data~\cite{talwar2020evaluating}. Therefore, comprehensive documentation and availability of these salient data characteristics are imperative for the reliable reproduction of the numerous \textbf{training} and \textbf{tensor} bugs in deep learning systems.

\textbf{Model:} A neural network model's architecture and implementation details are crucial to reproducing \textbf{model} and \textbf{API} bugs. The model architecture has 78.57\% and 70.00\% confidence values for model and API bugs, respectively, according to our generated rules. These high values signify that access to model details is vital for reliably reproducing issues stemming from model capacity, connectivity, and API usage.

An access to the model architecture (e.g., layer types, layer connectivity, weight initialization schemes, etc.) can help us systematically reproduce model bugs. Insufficient learning capacity in specific layers and improper weight initialization might lead to exploding/vanishing gradients~\cite{grosse2017lecture}, whereas incorrect layer connectivity might lead to representational bottlenecks~\cite{Tishby_Zaslavsky_2015}. These issues usually manifest as model bugs during training and inference. Thus, information about the model architecture can help us localize and reproduce such bugs. 

For reproducing the API bugs, model architecture can provide fundamental context. Details like layer dimensions, bottlenecks, parallelization needs, and memory requirements show how the model interacts with the API~\cite{yang2022comprehensive}. Bottlenecks within a model, areas where data processing slows down, and the need for parallel processing to handle large-scale computations shape how APIs are utilized. Additionally, the varying memory requirements of different architectures impact how the model leverages the system's resources via the API. This includes memory allocation for storing weights and activations and managing the flow of data through the network during operations like forward and backward propagation. Understanding the model architecture is vital for reproducing bugs, as it indicates whether issues stem from the model's design or from its interaction with the API. For example, a bug might arise due to the model's inability to handle certain operations efficiently, or it could be a result of the API not properly supporting specific architectural features. Hence, model architecture helps us reproduce the bugs triggered by incorrect usage of API. Therefore, the presence of model architecture allows the reproduction of both \textbf{model} bugs and \textbf{API} bugs in deep learning systems.

\textbf{Code Snippet:} Code snippets are invaluable for reproducing deep learning bugs since they isolate and encapsulate the core logic that triggers them. Developers can demonstrate and share the essence of buggy behaviour by creating a minimal reproducible example. Code snippets have very high confidence values of 86.0\% for training bugs, 72.41\% for tensor bugs, 71.43\% for model bugs, and 75.0\% for API bugs, according to our generated rules. These high values across all bug types signify that code snippets are critical for reliably reproducing deep learning bugs.

Each code snippet may contain the relevant model, data processing, training and evaluation scripts. Developers can use such a snippet to recreate the bug-inducing steps. For example, a snippet may compactly capture just a few lines, mishandling tensor shapes or performing incorrect gradient calculations, eliminating any confounding factors and enabling developers to reproduce these bugs. Developers can systematically execute, analyze, and debug the code to reproduce various bugs, including training, tensor, model, and API bugs.

\textbf{Logs:} Logs play a fundamental role in deep learning by facilitating the reproduction and resolution of bugs. They comprehensively record every detail during model training, capturing information about various conditions, configurations, and events before the failures. Logs have high confidence values of 76.00\% for training bugs, 93.10\% for tensor bugs, 78.57\% for model bugs, and 85.00\% for API bugs according to our generated rules. Such high confidence levels across all bug types highlight that comprehensive log recording is important for the reliable reproduction of deep-learning bugs.

These logs include essential components such as hyperparameters, dataset characteristics, hardware specifications, framework versions, and random number generator seed values. The true power of logs lies in their ability to recreate past training runs precisely. Developers can isolate and reproduce the environment that led to the original bugs using the logged hyperparameters, such as batch size, learning rate schedules, and gradient clipping thresholds. Logs can also help developers reproduce specific deep-learning bugs. For example, logged random seeds can help the developers recreate a specific weight initialization or data batching order, which can then be used to reproduce \textbf{training} bugs. Similarly, logs can be used to reproduce \textbf{Model}, \textbf{Tensor} and \textbf{API} Bugs. Thus, logs are invaluable for accurately recreating conditions that result in errors and enabling the reproduction of deep learning bugs.

\begin{table}[]
\centering
\captionsetup{justification=raggedright,singlelinecheck=false} 
\caption{Top 5 Edit Actions for Reproducing Specific Types of Deep Learning Bugs}
\label{tab:editOperationTypeOfBug}
\begin{tabular}{|l|l|}
\hline
\textbf{Training}    & \textbf{Model}        \\ \hline
\hline
Input Data Generation (0.5625)         & Hyperparameter Initialization (0.5142)       \\ \hline
Import Addition  (0.4583)              & Dataset Procurement (0.4390)              \\ \hline
Compiler Error Resolution (0.3750)      & Compiler Error Resolution (0.4146)   \\ \hline
Dataset Procurement (0.3542)            & Import Addition (0.4146)                         \\ \hline
Hyperparameter Initialization (0.3333)    & Neural Network Construction (0.3659)             \\ \hline
\hline
\textbf{Tensor}       & \textbf{API}          \\ \hline
\hline
Hyperparameter Initialization (0.5517)  & Input Data Generation (0.40)         \\ \hline
Input Data Generation (0.5172) & Hyperparameter Initialization (0.40) \\ \hline
Import Addition (0.5172) & Import Addition (0.35)            \\ \hline
Dataset Procurement (0.4138) & Logging (0.25)                       \\ \hline
Obsolete Parameter Removal (0.3448)  & Obsolete Parameter Removal (0.25)    \\ \hline
\end{tabular}
\end{table}
\subsubsection{Relationship between Edit Actions and Type of Bugs}
After determining association between the useful information and the type of bug, we derived the relationship between the edit actions and the type of bug. We use the Apriori algorithm to determine the relationship, as done earlier. Table.~\ref{tab:editOperationTypeOfBug} summarizes our findings, and we discuss them in detail below.

\textbf{Training Bug:} Input data generation (56.25\% confidence) is the most frequently used edit action for reproducing training bugs. However, our analysis (Table.~\ref{tab:criticalInformationStatistics}) shows that $\approx$77\% of the bug reports/SO posts contain information about the data. Leveraging this information and our proposed edit action -- Input Data Generation (A$_{1}$), we can reproduce the training bugs in deep learning systems. Furthermore, import addition (45.8\% confidence) and compiler error resolution (37.5\% confidence) are other edit actions which are often used to reproduce training bugs. Although 79\% of bug reports contain code snippets, they are often incomplete. To reproduce the bug effectively, we thus need to complete these snippets by adding the necessary import statements and migrating them to the latest versions of libraries and frameworks. Finally, dataset procurement (35.42\% confidence) and hyperparameter initialization (33.33\% confidence) may also help reproduce training bugs by procuring the dataset required and initializing the missing hyperparameters. 

Since the training bugs have specific sub-faults, we manually analyse the Stack Overflow posts and Github issue reports for different sub-types and discuss the edit actions, which can be used to reproduce the specific sub-faults below.

\begin{itemize}
\item \textbf{Optimisation:} Optimization bugs can be reproduced by employing a combination of edit actions, with a primary focus on three key areas: Hyperparameter Initialization (60.00\% confidence), Input Data Generation (60.00\% confidence), and Neural Network Construction (40.00\% confidence). These actions involve configuring the optimizer and learning rate, creating representative training data, and building the model architecture, respectively. Additionally, logging (40.00\% confidence) can help monitor the training process, while Import Addition (20.00\% confidence), Compiler Error Resolution (20.00\% confidence), and Version Migration (20.00\% confidence) ensure compatibility within the environment. By systematically testing different optimizer configurations, generating appropriate input data, and constructing the relevant model architecture, developers can effectively reproduce optimization issues and gain insight into the underlying causes.
\item \textbf{Loss Function:} Loss function bugs, which can arise from incorrect loss calculations or suboptimal loss function choices, can be reproduced through a combination of edit actions. Hyperparameter Initialization (66.67\% confidence) plays a crucial role in configuring the loss function while Logging (58.33\% confidence) captures the output of intermediate loss values during training. Since the design of the output layer in the neural network directly influences the calculation of loss, Neural Network Construction (41.67\% confidence) may be necessary to reproduce the bugs related to the loss function. Furthermore, it is essential to generate appropriate training data through Input Data Generation (33.33\% confidence) to effectively reproduce bugs related to the loss function.
\item \textbf{Hyperparameters:} Hyperparameter sub-faults, such as suboptimal batch size, suboptimal number of epochs, and suboptimal learning rate, can be reproduced by utilizing the Hyperparameter Initialization edit action (64.29\% confidence). This involves initializing various hyperparameters, such as batch size, epochs, and learning rate, that may be missing or suboptimal in the provided code. By experimenting with different commonly used values for these hyperparameters, we were able to reproduce them. In addition to that, Input Data Generation (50.00\% confidence) proves to be valuable for reproducing hyperparameter bugs. Generating representative input data is crucial for triggering hyperparameters-related issues during the training process. Furthermore, actions like Import Addition (42.86\% confidence), Logging (35.71\% confidence), and Compiler Error Resolution (35.71\% confidence) are helpful in ensuring that the code runs smoothly and allows for testing different hyperparameter configurations. These actions contribute to creating a compatible environment to address hyperparameter-related concerns effectively.
\end{itemize}

\textbf{Model Bug:} Model bugs are often reproduced by using the edit action -- hyperparameter initialization (51.4\% confidence). From our manual analysis, we observed that the hyperparameters were not often reported for model bugs. Hence, hyperparameter initialization was often used to reproduce the model bugs. Additionally, dataset procurement (43.90\% confidence) and compiler error resolution (41.46\% confidence) are typical edit actions to reproduce the model bugs. Moreover, the code snippets for deep learning bugs are often incomplete, as observed in our manual analysis; hence, import addition (41.46\% confidence) is a critical edit action for completing the code snippet and reproducing model bugs. Finally, model bugs might be reproduced by modifying a neural network's architecture, as shown by the moderate confidence value (36.6\%) for the edit action -- neural network construction. Since model bugs are primarily caused by errors in the neural network architecture~\cite{taxonomyRealFaults}, reconstructing the neural network might be the first step to reproduce them. Thus, model bugs can be reliably reproduced through several edits that initialize hyperparameters, resolve compiler errors, procure datasets, add imports, and construct neural networks. 

Similar to training bugs, model bugs also have specific sub-faults. Hence, we report the edit actions for reproducing specific sub-faults below.

\begin{itemize}
\item \textbf{Layer Type \& Properties:} To reproduce specific model bugs related to individual layers, such as incorrect layer types, suboptimal filter sizes, or inappropriate activation functions, it is necessary to carefully define the layers as part of the Neural Network Construction (78.57\% confidence) action. By accurately specifying the layers in the model architecture, developers can trigger the desired layer-specific bugs for further analysis and resolution. It is also crucial to initialize the appropriate hyperparameters through the Hyperparameter Initialization (57.14\% confidence) action. Additionally, obtaining the required input data shapes via Input Data Generation (42.86\% confidence) is essential for triggering layer-specific issues during the training process. To ensure compatibility within the project environment, actions such as Compiler Error Resolution (42.86\% confidence) and Version Migration (42.86\% confidence) are beneficial for updating the layer definitions and reproducing any compatibility issues that may arise. 
\item \textbf{Model Type \& Properties:} Bugs associated with suboptimal model architecture, incorrect network structure, or missing layers can often be reproduced using the Neural Network Construction edit action (71.43\% confidence). By carefully constructing the neural network based on the architecture details provided in a bug report, developers can reproduce sub-faults from the model bugs category. The interaction between the hyperparameters and the model architecture could be important for model bugs. Therefore, the Hyperparameter Initialization action (42.86\% confidence) also plays a significant role in effectively reproducing model-related issues. To ensure compatibility with the latest library and framework version, actions such as Import Addition (35.71\% confidence), Obsolete Parameter Removal (28.57\% confidence), and Compiler Error Resolution (28.57\% confidence) are essential. These actions help update the codebase and ensure that the necessary dependencies are met to define the desired model architecture accurately.
\end{itemize}

\textbf{Tensor Bug:} Data is the most important information for reproducing tensor bugs, as shown by the strong correlation between Tensor Bug \& Data (see Table.~\ref{tab:criticalInformationBugType}). This phenomenon can be attributed to the fact that tensor bugs primarily relate to the data~\cite{taxonomyRealFaults}. Since tensor bugs are so data-dependent, dataset procurement emerges as an important edit action to reproduce them. Bug reports reference the exact dataset used (if well-known) or describe the data type, shape and distribution. With this information, input data generation and dataset procurement can generate or procure the data required for the reproduction of the tensor bugs. Like other deep learning bugs, import addition (51.72\% confidence) and hyperparameter initialization (55.17\% confidence) also play a key role in completing the code snippet and reproducing the tensor bugs. Thus, tensor bugs can be reliably reproduced through several edits that procure the datasets, generate input data, add logging, import the required dependencies, and initialize the hyperparameters.

\textbf{API Bug:} Since API Bugs stem from an improper usage of application programming interfaces (APIs), they do not focus as heavily on the model training process. As a result, bug reports might lack detailed information about the input data or hyperparameters used. This encourages the use of common edit actions such as input data generation (40.00\% confidence) and hyperparameter initialization (40.00\% confidence) when reproducing API bugs. Additionally, code snippets in bug reports are often incomplete, necessitating edit actions like import addition (35.00\% confidence) and logging (25.00\% confidence) to trace the intermediate states and outputs of the API. Thus, API bugs can be reliably reproduced through several edit actions that generate input data, import the required dependencies, add logging, initialize hyperparameters and remove obsolete parameters. 
\begin{shaded}
\noindent \textbf{Summary of RQ2:} By applying the Apriori algorithm on the data produced from our reproduction of  \textbf{148} deep learning bugs, we identified the \textbf{top 3} most important pieces of information and the \textbf{top 5} edit actions needed to reproduce each category of deep learning bug. This provides insights into the missing information that should be solicited in bug reports as well as the edit actions required to reproduce specific bug types.
\end{shaded}
\subsection{\textit{RQ3: How do the suggested edit actions and component information affect the reproducibility of deep learning bugs?}}

We conduct a developer study to determine if our suggested edit actions and information help one reproduce deep learning bugs. We prepared four sets of bugs, each consisting of two different types of bugs (see Section 4.6). We randomly assigned one of the four bug sets to each participant and instructed them to reproduce the bugs in their assigned set. We divided the participants into the control and experimental groups using stratified random sampling (see Section 4.6). The developers in both the control and experimental groups reproduced the same bugs from our prepared set of bugs. The only difference was that the experimental group received hints based on our findings to help reproduce the bugs, while the control group did not receive these hints. Since both groups received randomly selected sets from the same pool of four identical bug sets, each bug was reproduced independently by at least five developers across both control and experimental groups.

\begin{figure}
    \centering
    \includegraphics[width=\linewidth]{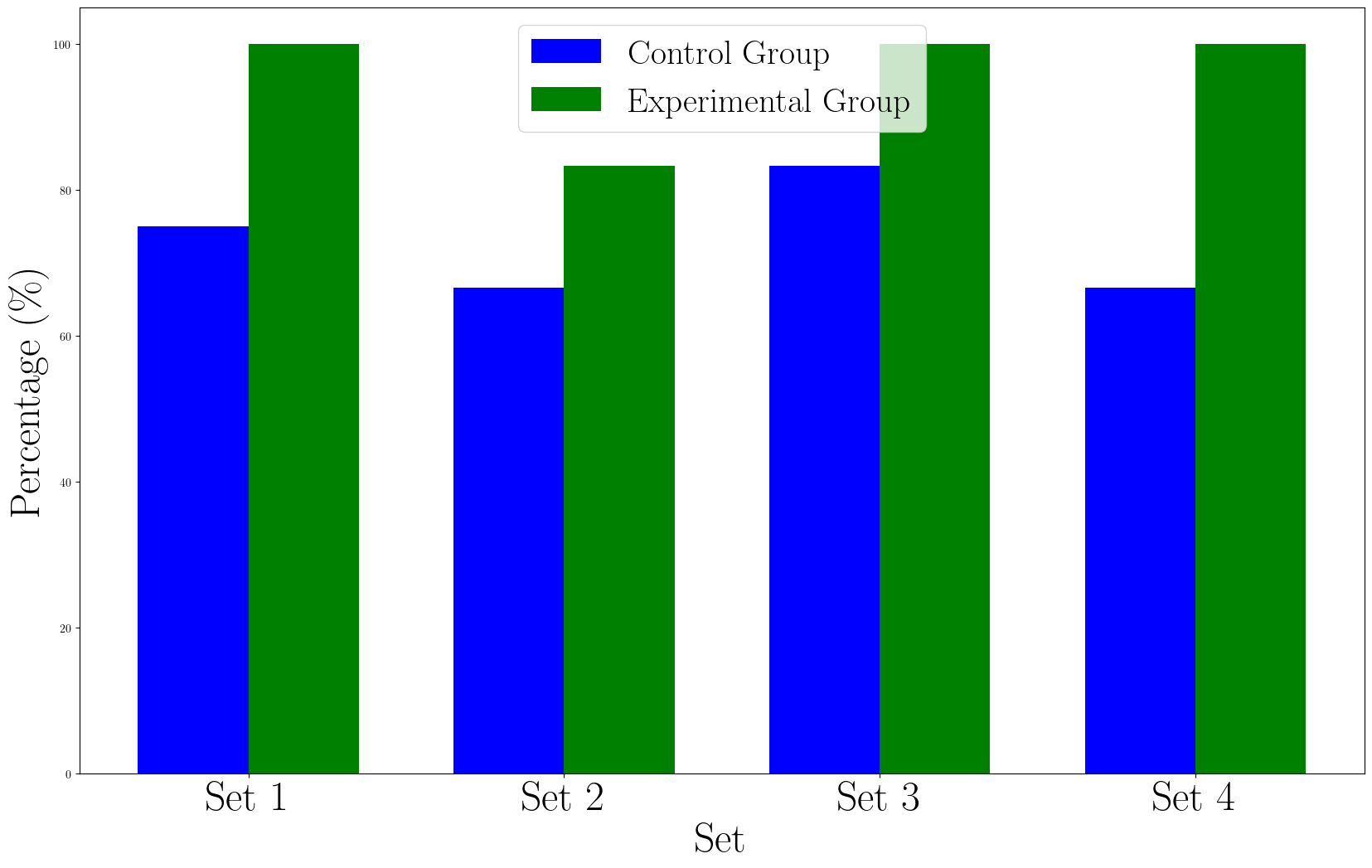}
    \caption{Bug Reproduction Success Rates by Control and Experimental Groups}
    \label{fig:BugReproductionSuccessRates}
\end{figure}

Table.~\ref{tab:reproducibilityRate} and Fig.~\ref{fig:BugReproductionSuccessRates} shows the bug reproducibility rate of the control and experimental groups across different sets of bugs. We found that developers in the control group could reproduce \textbf{72.91\%} bugs without hints. In contrast, the developers in the experimental group could reproduce \textbf{95.83\%} of bugs with our hints - a \textbf{22.92\%} increase. This significant increase in reproducibility rate demonstrates that our identified edit actions and information improved developers' ability to reproduce deep learning bugs. 
\begin{table}[]
\caption{Percentage of bugs successfully reproduced by control and experimental group across different sets.}
\centering
\resizebox{\columnwidth}{!}{%
\begin{tabular}{|p{0.15\linewidth}|p{0.35\linewidth}|p{0.35\linewidth}|p{0.15\linewidth}|}
\hline
 & \textbf{\% of bugs successfully reproduced by Control group} & \textbf{\% of bugs successfully reproduced by Experimental group} & {\% Increase}\\ \hline
 \hline
\textbf{Set 1} & 75.00 & 100.00 & 25.00 \\ \hline
\textbf{Set 2} & 66.66 & 83.33 & 16.66 \\ \hline
\textbf{Set 3} & 83.33 & 100.00 & 16.66 \\ \hline
\textbf{Set 4} & 66.66 & 100.00 & 33.33 \\ \hline
\textbf{Average} & \textbf{72.91} & \textbf{95.83} & \textbf{22.92} \\ \hline
\end{tabular}%
}
\label{tab:reproducibilityRate}
\end{table}

We also collect the qualitative feedback from the participants to better understand their pain points in bug reproduction. Their feedback provided new insights not covered in our original study. Notably, \textbf{31.78\%} of developers highlighted the \textit{lack of standardized debugging tools} as the primary challenge in bug reproduction. Furthermore, the developers mentioned \textit{missing information (data, logs, hyperparameters, and dependencies)}, and \textit{lack of unit testing or version control in DL systems} as the most challenging aspects of reproducing deep learning bugs, as shown below.

\begin{shaded}

\noindent \textbf{Question}: What are the challenges of bug reproduction in day-to-day activities?

\noindent \textit{R1:} lack of debugging support and missing information

\noindent \textit{R2:} data quality issues and lack of standardized debugging procedures and tools.

\noindent \textit{R3:} no standardized debugging practices, and lack of clarity on the information needed, also lot of dependencies (libraries, framework, data, infra and so on)

\noindent \textit{R4:} the flaky nature of deep learning models, and the unclear expectations of how model is supposed to behave.

\noindent \textit{R5:} memory issues, documentation issues (missing information in issues), weak debugging support

\noindent \textit{R6:} version control of deep learning is tricky, because of multiple snapshots of models, model management and reproducibility is tricky. also, the lack of standardized debugging practices makes it more tricky.

\noindent \textit{R7}: distributed computing makes it difficult to find and reproduce the bugs. test coverage is also a problem as we cannot find the bugs properly because of lack of coverage and that is a problem with the reproduction of bugs.
\end{shaded}

We also analyze how our suggested actions and information help the participants in bug reproduction. According to the qualitative responses, \textbf{40.91\%} of developers found our suggested edit actions to be helpful for reproducing their assigned bugs. \textbf{54.55\%} of developers report that our suggested hints about the useful information helped them narrow down where to look in the code. Thus, the qualitative feedback below highlights the benefits of our findings in real-life settings.

\begin{shaded}
\noindent \textit{R1}: I followed the guidance in the survey and used the hints to generate the appropriate training dataset, which then allowed me to reliably reproduce the problematic behavior during model training.

\noindent \textit{R2}: had to add manual imports for torch and nn, and fixed the compiler errors related to imports, as highlighted by the hint provided

\noindent \textit{R3}: Following the hint, I systematically generated all of the necessary input data that would be required in order to reliably reproduce the software bug during testing.

\noindent \textit{R4}: The hint mentions the Iris dataset, so I used the edit action called ``Dataset Procurement" and got the dataset, downloaded it and edited the code snippet to reproduce the bug.

\noindent \textit{R5}: with the given hint, I generated the input data required for the bug reproduction.

\noindent \textit{R6}: The imports were missing and as per the hint, data frame was generated for specific columns, which helped the resolution of the bug.

\noindent \textit{R7}: As hinted in the survey, explicitly specifying the columns when constructing the data frame helped in the bug reproduction
\end{shaded}

\begin{figure}
    \centering
    \includegraphics[width=1\linewidth]{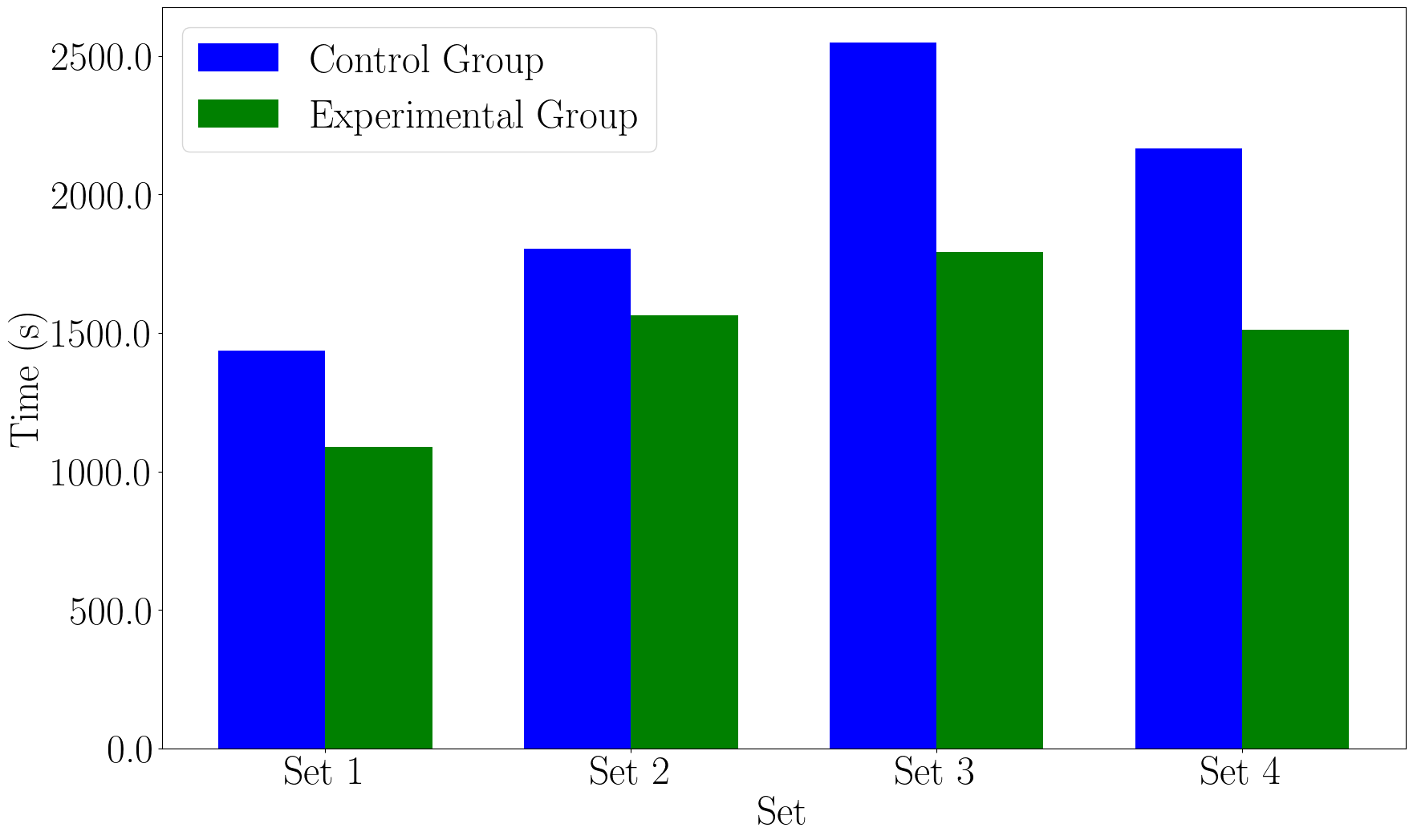}
    \caption{Time Taken for Bug Reproduction Across Control and Experimental Groups}
    \label{fig:timeForBugReproduction}
\end{figure}

We also calculate the time the control and experimental groups took to reproduce their assigned bugs. This helps us assess our findings' benefits in reducing the time required for deep learning bug reproduction. Table.~\ref{tab:timeTakenForReproduction} and Fig.~\ref{fig:timeForBugReproduction} show the time taken to reproduce bugs by the control and experimental groups. The experimental group outpaced the control group in bug reproduction across all sets, with the most notable difference in Set 4, where the experimental group was \textbf{30.16\%} faster. On average, the control group lagged behind the experimental group by \textbf{24.35\%} across all sets. These results demonstrate that our recommended edit actions and \textit{component information} enabled the experimental group to reproduce deep learning bugs much faster than the control group that did not receive this information.
\begin{table}[]
\caption{Time taken for bug reproduction by control and experimental groups}
\centering
\resizebox{\columnwidth}{!}{%
\begin{tabular}{|p{0.15\linewidth}|p{0.35\linewidth}|p{0.35\linewidth}|p{0.15\linewidth}|}
\hline
 & \textbf{Average time taken by Control Group (seconds)} & \textbf{Average time taken by Experimental Group (seconds)} & \textbf{\% Decrease} \\ \hline
 \hline
\textbf{Set 1} & 1437 & 1088 & \textbf{24.28} \\ \hline
\textbf{Set 2} & 1803 & 1563 & \textbf{13.31} \\ \hline
\textbf{Set 3} & 2548 & 1792 & \textbf{29.67} \\ \hline
\textbf{Set 4} & 2165 & 1512 & \textbf{30.16} \\ \hline
\end{tabular}%
}
\label{tab:timeTakenForReproduction}
\end{table}

\subsubsection{Impact of Hints on Bug Reproducibility}
To determine the impact of our hints on bug reproducibility, we constructed a generalized linear model (GLM) using the Binomial family with a logit link function~\cite{nelder1972generalized}. This allowed us to test the statistical significance of multiple independent factors on bug reproducibility (i.e, our dependent variable). The factors were experience with deep learning bug fixing, experience with deep learning, profession, and the presence/absence of hints, with `Hints' being our main factor of interest. The DLBugFixExp\_* factors represent the participants' experience in fixing deep learning bugs, with levels 0, 1, and 2 corresponding to experience of 0-4, 5-9, and 10+ years of experience, respectively. We chose Odds Ratio as our effect size metric for two main reasons. First, it has been commonly used in similar studies within software engineering research, as demonstrated by Ceccato et al.~\cite{ceccato2014family}. Second, the Odds Ratio is appropriate for a logistic regression-based model with a binary target variable, which we used in our study.

\begin{table}[h]
\centering
\caption{GLM Model for Assessing the Impact of Various Factors on the Reproducibility of Deep Learning Bugs}
\label{tab:regression}
\begin{tabular}{lccccc}
\toprule
Variable & Estimate & Std. Error & z value & Pr $>$ $|$z$|$ & Effect Size (OR) \\
\midrule
Intercept & -3.4229 & 1.898 & -1.803 & 0.071 & - \\
DLBugFixExp\_0 & 0.2251 & 0.856 & 0.263 & 0.793 & 1.252493 \\
DLBugFixExp\_1 & -0.2449 & 0.564 & -0.435 & 0.664 & 0.782786 \\
DLBugFixExp\_2 & -3.4032 & 2.283 & -1.491 & 0.136 & 0.033268 \\
Hints & 3.0899 & 0.991 & 3.118 & \textbf{0.002} & 21.974308 \\
DLExp & 2.5448 & 1.725 & 1.475 & 0.140 & 12.740844 \\
Field & 0.1915 & 1.145 & 0.167 & 0.867 & 1.211112 \\
\bottomrule
\end{tabular}
\end{table}

Based on the regression results in Table \ref{tab:regression}, we can see that the presence of hints had a statistically significant positive effect on the reproducibility of deep learning bugs (p = 0.002 $<$ .05). The effect size, as measured by the odds ratio, indicates that the presence of hints increases the odds of reproducing a deep learning bug by a factor of 21.97 compared to the absence of hints. The intercept term, with an estimate of -3.4229 (p = 0.071), represents the baseline probability of reproducing a deep learning bug when no hints are provided, the participant has no experience with deep learning bug fixing, deep learning in general, and their profession is not considered. Moreover, as shown in Table.~\ref{tab:regression}, factors like experience with deep learning bug fixing, general deep learning experience, and profession (academia vs industry) did not significantly influence reproducibility.

Overall, the above results suggest that the presence of targeted hints positively impacts the reproducibility of deep learning bugs with a statistically significant margin and a large effect size. Other factors, such as experience and profession, do not play a significant role in bug reproducibility.

\begin{shaded}
\noindent \textbf{Summary of RQ3:} In the user study, developers were assigned to either the experimental or control group where the experimental group received our suggested edit actions and component information. The experimental group \textbf{reproduced 22.92\% more deep learning bugs} and \textbf{decreased reproduction time by 24.35\%} compared to the control group. This demonstrates that the identified edit actions and component information substantially improve the reproducibility rate and reduce the time needed to reproduce deep learning bugs.
\end{shaded}
\section{Discussions} \label{section6}
In this section, we first provide actionable insights about the reproducibility of deep learning bugs and recommend potential directions for further research (see Section 6.1). We then demonstrate how our findings can improve the reproducibility of the DL bugs with the use of large language models (e.g., Llama 3) (see Section 6.2).

\subsection{\textit{Reproducibility of Deep Learning Bugs}}

From our manual reproduction and user study, we observe that \textit{API}, \textit{Model}, and \textit{Tensor} bugs are relatively more straightforward to reproduce. This behaviour can be explained by the fact that these bugs are more specific to the location in the code where they originate. For example, faulty input data usually triggers tensor bugs, whereas incorrect usage of a framework's API causes API bugs. In contrast, training bugs cover multiple issues related to deep learning model training, and GPU bugs relate to GPU devices for deep learning and manifest across GPU interactions. 

This behaviour is further supported by the reproducibility rates and efforts involved in reproducing different types of bugs. The reproducibility rates of Training bugs and GPU bugs were 89.65\% and 42.85\%, respectively. On the other hand, the reproducibility rates for API, Model and Tensor bugs were 81.81\%, 88.46\%, and 80.75\%, respectively. Even though the training bugs could be reproduced reliably, the efforts involved in reproducing the training bugs were significantly more than those of other bugs. The average time to reproduce the API, Model and Tensor bugs was 45.5, 43.9 and 43.8 minutes, respectively. On the other hand, the average time to reproduce the Training bugs was 52.5 and 57.33 minutes, respectively. These statistics highlight that we need to put more effort into reproducing the Training and GPU bugs, which calls for further research into the specific nature of Training and GPU bugs. 

To assist future research in the reproducibility of deep learning bugs, we provide directions for future research below:

\textit{Understanding Training Bug Reproducibility}: Training bugs are the most common type of bug in deep learning systems, accounting for 52.5\% of all bugs~\cite{taxonomyRealFaults}. Training bugs have high reproducibility rates but also require significant effort to reproduce. This behaviour presents an opportunity to understand better what factors make training bugs more reproducible or harder to reproduce. By analyzing the training procedures, model architectures, optimizers, hyperparameters, and other elements that either aid or hinder reproducibility, we can uncover insights to guide the diagnosis and repair of training bugs. There is also room to develop improved tools and methodologies explicitly focused on efficiently reproducing the nuanced nature of training bugs.

\textit{Analyzing the GPU Bug Reproducibility Gap}: The lower reproducibility rate for GPU bugs highlights a gap in understanding the interactions between deep learning code and the underlying GPU hardware or drivers. By further studying irreproducible GPU bugs and quantifying the aspects that impede reproducibility, such as hardware differences, software dependencies, and environmental factors, we can work towards solutions to increase reproducibility. Opportunities exist to build infrastructure, leverage containerization or virtualization, and create testing tools to better control for sources of non-determinism that impact GPU bug reproduction.

\textit{Reproducible Testbeds for Deep Learning:} Ultimately, the variability in reproducing different categories of deep learning bugs motivates the need for reproducible testbeds. Shared sets of reproducible deep learning bugs with associated test cases, model architectures, training configurations, dependencies, and environmental contexts will accelerate future research. Constructing such testbeds requires systematic characterization of how these factors influence reproducibility. Reproducible testbeds also support developing specialized techniques for efficiently reproducing and debugging deep learning bugs.

\subsection{\textit{Challenges in Reproducing Deep Learning Bugs: A Comparison between Stack Overflow and GitHub}}
Reproducing deep learning bugs from Stack Overflow posts and GitHub repositories presents different challenges and requires varying levels of effort. The main differences lie in the availability of code and data, environmental setup, context, completeness, and reproducibility expectations. We explain them briefly as follows.

First, GitHub repositories often provide the complete codebase and sometimes the accompanying datasets, whereas Stack Overflow posts typically include small code snippets without the broader context and data. Hence, we need more effort in reconstructing the code and generating synthetic data when reproducing bugs from Stack Overflow. Second, GitHub repositories commonly include setup instructions, dependency files, and development environment configurations, making it easier to recreate the original environment accurately. On the contrary, Stack Overflow posts rarely provide such details, requiring guesswork about software versions, dependencies, and environment settings. Finally, GitHub issues tend to have more contextual information, such as detailed problem descriptions, steps to reproduce a bug, and error logs supporting root cause analysis of the issue. Stack Overflow posts may lack this level of detail, making it harder to comprehend and reproduce a bug accurately.

Despite the additional resources available in GitHub repositories, reproducing bugs from them still presents several challenges:
\begin{itemize}
    \item \textit{Incomplete or outdated repositories}: Critical files, dependencies, or code changes may be missing or outdated, making it difficult to reproduce the exact environment and conditions under which a bug occurred.
    \item \textit{Large and complex codebases}: Deep learning projects often have extensive and intricate codebases, requiring significant time and expertise to set up and navigate them.
    \item \textit{Proprietary or sensitive data}: Many projects may involve proprietary data that cannot be shared publicly, making it challenging to reproduce data-related bugs without access to the original data.
    \item \textit{Insufficient documentation}: Many GitHub repositories fail to provide clear documentation on project architecture, installation steps, and expected behaviours. The lack of clear documentation makes it difficult for developers to reproduce bugs, as they face challenges in understanding the codebase and configuring the project locally.
    \item \textit{Dependency management}: Resolving dependency conflicts or managing compatibility issues across different project versions can be a significant challenge when reproducing bugs.
\end{itemize}

To address these challenges, we recommend several changes or adaptations to the status quo as follows. First, Stack Overflow should leverage its user base and search functionality for discussing common deep-learning bugs and quick solutions, while GitHub should be the primary platform for in-depth bug reproduction and resolution. Second, Stack Overflow should introduce a specialized format for deep learning bug reports, expanding on its traditional minimum working example approach to include dataset details, framework versions, hardware specifications, and reproducible code snippets. GitHub repositories should be kept up-to-date, well-documented, and complete. Finally, both platforms should incentivize active participation in bug reporting and resolution through specialized badges, reputation points, or recognition. Stack Overflow could introduce ``Deep Learning Debug Rooms" for real-time, collaborative problem-solving, and GitHub could highlight top contributors in bug resolution. By combining Stack Overflow's community-driven approach with GitHub's comprehensive project management features, we can develop a more efficient ecosystem for deep learning bug resolution, leading to more robust deep learning systems.

\section {Threats to Validity} \label{section7}
Threats to \textit{external validity} relate to the generalizability of our findings. We have reproduced multiple bugs from five different types to mitigate this threat. Moreover, we have reproduced bugs from 14 different architectures, ranging from Logistic Regression, CNN, Transformers and BERT-based architectures. Finally, we have also reproduced the bugs from the past six years (2017-2023), and our findings align with and extend the previous findings of software bug reproducibility~\cite{msr2019mondal}, possibly indicating the generalizability of our study.

Threats to \textit{internal validity} pertain to experimental errors and confounding variables during the reproduction of bugs. The possibility of introducing new bugs into the code exists during bug reproduction. We have implemented several strategies and precautions to mitigate these threats in our experimental design. For instance, bug reproduction attempts are carried out in controlled and isolated Python virtual environments. We refrain from using hardcoded or fixed values in our code snippets to prevent potential errors during bug reproduction. For instance, during the bug reproduction, we utilize random hyperparameter initialization, construct multiple neural networks, and generate randomized input data that aligns with the original distribution. This methodology mitigates the influence of unforeseen variables on the outcomes of our experiments. To further mitigate this threat, we ran the updated code snippets five times for all the edit actions. Following bug reproduction, we undertake rigorous manual analysis to identify any discrepancies or anomalies in the outcomes. 

A potential threat to the validity of our findings arises from the inclusion of context-specific, natural language hints from bug reports. This could potentially introduce bias when assessing the experimental group's performance. However, our analysis of qualitative feedback from participants indicates that participants primarily benefited from the hints derived from our Apriori algorithm (RQ2), with minimal mention of context-specific hints from the bug report. In particular, the participants reported using the guidance to generate appropriate datasets, address import issues, create necessary input data, and apply specific edit actions, which are closely tied to our algorithm's outputs rather than the natural language hints. Thus, the threat posed by the inclusion of natural language hints might be negligible.

Another threat to validity is the incomplete or incorrect information in the bug report,  which could impact the applicability of certain edit actions suggested by our technique. In particular, edit actions that require additional context from the report, such as those involving input data generation, neural network construction, dataset procurement, and downloading models/tokenizers, may be limited by missing details in the reports. However, the majority of our suggested edit actions focus on code modifications that can be derived from the provided code snippets without relying heavily on external context. Therefore, while missing information may constrain our capability for data and model-driven actions, our core technique and findings remain widely applicable for many code-focused edit recommendations. To mitigate this threat further, we apply extensive filtration based on the criteria suggested by Moravati et al.~\cite{defects4ml} and Humbatova et al.~\cite{taxonomyRealFaults}, which leads to a clean and generalizable dataset.

Finally, the way in which the contextual information is presented to the developers could potentially threaten the validity of the control and experiment groups in the user study. To mitigate this threat, we ensure that the contextual information is presented in an identical format to the control and experiment groups.

Threats to \textit{construct validity} relate to the use of appropriate metrics for evaluating the results of a user study. In the user study, we used the reproducibility rate and time to reproduce the deep learning bugs as the metrics to evaluate the benefits of our findings. To measure these accurately, we had participants justify why and how they used the edit actions or found certain information important. We verified these justifications against our ground truth to determine if the bug was successfully reproduced. After checking the user justifications, we calculated the reproducibility rate and spent time. By including user justifications and verifying them, we ensured that the metrics directly measured how easily bugs could be reproduced. We also had clear criteria for determining reproduction success and used Opinio for precise timing data~\cite{Opinio}, mitigating threats related to measurement errors or subjective assessments. 

Another threat to the validity might be the accuracy of tags used to categorize the bugs. The incorrect categorization of bugs might lead to incorrect conclusions for different bugs. To mitigate this threat, we only select the tags that are present in the taxonomy and the sub-taxonomies defined by Humbatova et al.~\cite{taxonomyRealFaults}. For example, we specifically chose the ``loss-function" as one of the tags to distinguish the training bugs since the Loss Function falls within the sub-taxonomy of Training bugs. Thus, the threats to construct validity were effectively mitigated.

\section {Related Work} \label{section8}

There have been several studies that focus on the aspects of reproducibility of software bugs~\cite{msr2019mondal, mondal2022reproducibility, mondalreproducibility}, or focus on why some bugs cannot be reproduced~\cite{icsme2020rahman, rahman2022works}. Many studies attempt to learn the nature of deep learning bugs~\cite{tfProgramBugs, islamfse19, taxonomyRealFaults, tarek2022icsme}, and how they can be localized automatically~\cite{nerdbug, yan2021exposing, zhang2020detecting}. A few studies also attempt to learn about the state of reproducibility of deep learning in software engineering~\cite{trainingReproducibleDL,reproducibilityDLTSE}, and reproduce deep learning bugs as a part of benchmark dataset creation~\cite{defects4ml}. Unfortunately, only a little research has been done to understand the challenges in reproducing deep learning bugs and how we can improve their reproducibility.

Mondal et al.~\cite{msr2019mondal} extensively investigated the reproducibility of programming issues reported on Stack Overflow. They identified several key edit actions required to reproduce programming errors and traditional software bugs. While their work is a source of inspiration, it does not provide the edit actions for reproducing deep learning bugs. Since their dataset is constructed in Java programming language with no questions related to deep learning, their findings might not apply to deep learning bugs.

Chen et al.~\cite{trainingReproducibleDL} proposed a unified framework to train reproducible deep learning models. They also provide reproducibility guidelines and mitigation strategies for conducting a reproducible training process for DL models. However, the primary focus of their study was on the reproducibility of deep learning models, not deep learning bugs. Furthermore, the guidelines for deep learning models might not always extend to deep learning bugs. 

Moravati et al.~\cite{defects4ml} constructed the faultload benchmark containing deep learning bugs, \textit{defects4ML}. As a part of their benchmark dataset, they reproduced 100 bugs as a part of the reproducibility criterion for benchmark datasets. While their study was the first to reproduce deep learning bugs actively, they did not report the techniques and actions used to reproduce bugs. Furthermore, they did not report the useful information from the issue reports, which helped them reproduce the bugs.

Unlike many earlier studies above, we conduct an extensive empirical study to understand the challenges of deep learning bug reproduction and how we can improve the reproducibility of deep learning bugs. We construct a dataset of 668 bugs and manually reproduce 148 bugs, spanning three frameworks, five types of bugs, and 14 architectures. We not only (1) define ten edit actions which can be used to reproduce deep learning bugs but also (2) explore the associations among the type of bugs, the information required to reproduce them, and the specific edit actions that can be used to reproduce them, which makes our work novel. Our findings are also generalizable to Tensorflow and PyTorch bugs due to the diversity of our dataset. To ensure transparency and reproducibility, we have made our dataset and replication package publicly available\footnote{https://github.com/mehilshah/Bug\_Reproducibility\_DL\_Bugs}.

\section {Conclusion \& Future Work} \label{section9}
The reproducibility of deep learning bugs is a significant challenge for software practitioners, given that these bugs can lead to severe consequences. In this paper, we conduct an empirical study by reproducing 148 deep learning bugs and identifying ten key edit actions that could reproduce these bugs. Furthermore, we investigate the relationships among bug types, component information, and edit actions. To validate our findings, we conducted a user study with 22 participants. The developers equipped with our recommended edit actions and information could reproduce 22.92\% more bugs using 24.35\% less time compared to the developers without such hints. This demonstrates the real-world value of our findings in improving the reproducibility of deep learning bugs. Finally, we provide actionable insights about the state of deep learning bugs that could improve the reproducibility of deep learning bugs. To demonstrate the practical use of our findings, we show how they aid large language models in improving the reproducibility of deep-learning bugs. For future work, we plan to expand our study to additional deep-learning frameworks and model types. Moreover, we plan to conduct a large-scale study to determine the sufficiency of information in deep learning bug reports, similar to the study conducted by Mondal et al.~\cite{mondal2024can}. Using these insights, we will build a tool that automatically completes the reports for better bug reproducibility. Furthermore, our findings open up the ability to create automated systems for reproducing deep learning bugs. By training machine learning models on our dataset of reproduction actions matched to bug reports, we can build AI agents that take bug reports as input and automatically generate reproducibility scripts, similar to the work by White et al.~\cite{reproducibleBugReports}. This would significantly reduce the manual effort required to reproduce bugs and enable large-scale debugging and testing of deep learning systems.

\section*{Data Availability Statement (DAS)}
All the data generated or analyzed during this study are available in the GitHub repository to help reproduce our results~\cite{replicationPackage}.
\section*{Conflict of Interest}
The authors declare that they have no conflict of interest.

\section*{Acknowledgements}
We would like to thank Usmi Mukherjee for her invaluable contributions during the manual bug reproduction process as an independent collaborator. 

\bibliographystyle{IEEEtran}
\bibliography{bibliography}

\begin{thebibliography}{10}
\providecommand{\url}[1]{#1}
\csname url@samestyle\endcsname
\providecommand{\newblock}{\relax}
\providecommand{\bibinfo}[2]{#2}
\providecommand{\BIBentrySTDinterwordspacing}{\spaceskip=0pt\relax}
\providecommand{\BIBentryALTinterwordstretchfactor}{4}
\providecommand{\BIBentryALTinterwordspacing}{\spaceskip=\fontdimen2\font plus
\BIBentryALTinterwordstretchfactor\fontdimen3\font minus \fontdimen4\font\relax}
\providecommand{\BIBforeignlanguage}[2]{{%
\expandafter\ifx\csname l@#1\endcsname\relax
\typeout{** WARNING: IEEEtran.bst: No hyphenation pattern has been}%
\typeout{** loaded for the language `#1'. Using the pattern for}%
\typeout{** the default language instead.}%
\else
\language=\csname l@#1\endcsname
\fi
#2}}
\providecommand{\BIBdecl}{\relax}
\BIBdecl

\bibitem{med1}
D.~Shen, G.~Wu, and H.-I. Suk, ``Deep learning in medical image analysis,'' \emph{Annual review of biomedical engineering}, vol.~19, pp. 221--248, 2017.

\bibitem{fin1}
P.~M. Addo, D.~Guegan, and B.~Hassani, ``Credit risk analysis using machine and deep learning models,'' \emph{Risks}, vol.~6, no.~2, p.~38, 2018.

\bibitem{cyb1}
D.~S. Berman, A.~L. Buczak, J.~S. Chavis, and C.~L. Corbett, ``A survey of deep learning methods for cyber security,'' \emph{Information}, vol.~10, no.~4, p. 122, 2019.

\bibitem{pei2017}
\BIBentryALTinterwordspacing
K.~Pei, Y.~Cao, J.~Yang, and S.~Jana, ``Deepxplore: Automated whitebox testing of deep learning systems,'' \emph{Commun. ACM}, vol.~62, no.~11, p. 137–145, oct 2019. [Online]. Available: \url{https://doi.org/10.1145/3361566}
\BIBentrySTDinterwordspacing

\bibitem{ma2018}
\BIBentryALTinterwordspacing
L.~Ma, F.~Juefei-Xu, F.~Zhang, J.~Sun, M.~Xue, B.~Li, C.~Chen, T.~Su, L.~Li, Y.~Liu, J.~Zhao, and Y.~Wang, ``Deepgauge: Multi-granularity testing criteria for deep learning systems,'' in \emph{Proceedings of the 33rd ACM/IEEE International Conference on Automated Software Engineering}, ser. ASE '18.\hskip 1em plus 0.5em minus 0.4em\relax New York, NY, USA: Association for Computing Machinery, 2018, p. 120–131. [Online]. Available: \url{https://doi.org/10.1145/3238147.3238202}
\BIBentrySTDinterwordspacing

\bibitem{esteva2019guide}
A.~Esteva, A.~Robicquet, B.~Ramsundar, V.~Kuleshov, M.~DePristo, K.~Chou, C.~Cui, G.~Corrado, S.~Thrun, and J.~Dean, ``A guide to deep learning in healthcare,'' \emph{Nature medicine}, vol.~25, no.~1, pp. 24--29, 2019.

\bibitem{BRAIEK2020110542}
\BIBentryALTinterwordspacing
H.~B. Braiek and F.~Khomh, ``On testing machine learning programs,'' \emph{Journal of Systems and Software}, vol. 164, p. 110542, 2020. [Online]. Available: \url{https://www.sciencedirect.com/science/article/pii/S0164121220300248}
\BIBentrySTDinterwordspacing

\bibitem{islamfse19}
\BIBentryALTinterwordspacing
M.~J. Islam, G.~Nguyen, R.~Pan, and H.~Rajan, ``A comprehensive study on deep learning bug characteristics,'' ser. ESEC/FSE 2019.\hskip 1em plus 0.5em minus 0.4em\relax New York, NY, USA: Association for Computing Machinery, 2019, p. 510–520. [Online]. Available: \url{https://doi.org/10.1145/3338906.3338955}
\BIBentrySTDinterwordspacing

\bibitem{selfdrivingcarcrash}
\BIBentryALTinterwordspacing
D.~Wakabayashi, ``Self-driving uber car kills pedestrian in arizona, where robots roam,'' Mar 2018, accessed on December 17, 2023. [Online]. Available: \url{https://www.nytimes.com/2018/03/19/technology/uber-driverless-fatality.html}
\BIBentrySTDinterwordspacing

\bibitem{nagarajan2018impact}
P.~Nagarajan, G.~Warnell, and P.~Stone, ``The impact of nondeterminism on reproducibility in deep reinforcement learning,'' 2018.

\bibitem{krishnan2020against}
M.~Krishnan, ``Against interpretability: a critical examination of the interpretability problem in machine learning,'' \emph{Philosophy \& Technology}, vol.~33, no.~3, pp. 487--502, 2020.

\bibitem{defects4ml}
\BIBentryALTinterwordspacing
M.~M. Morovati, A.~Nikanjam, F.~Khomh, and Z.~M.~J. Jiang, ``Bugs in machine learning-based systems: A faultload benchmark,'' \emph{Empirical Softw. Engg.}, vol.~28, no.~3, apr 2023. [Online]. Available: \url{https://doi.org/10.1007/s10664-023-10291-1}
\BIBentrySTDinterwordspacing

\bibitem{msr2019mondal}
S.~Mondal, M.~M. Rahman, and C.~K. Roy, ``Can issues reported at stack overflow questions be reproduced? an exploratory study,'' in \emph{2019 IEEE/ACM 16th International Conference on Mining Software Repositories (MSR)}, 2019, pp. 479--489.

\bibitem{icsme2020rahman}
M.~M. Rahman, F.~Khomh, and M.~Castelluccio, ``Why are some bugs non-reproducible? : –an empirical investigation using data fusion–,'' in \emph{2020 IEEE International Conference on Software Maintenance and Evolution (ICSME)}, 2020, pp. 605--616.

\bibitem{liang2022gdefects4dl}
Y.~Liang, Y.~Lin, X.~Song, J.~Sun, Z.~Feng, and J.~S. Dong, ``gdefects4dl: a dataset of general real-world deep learning program defects,'' in \emph{Proceedings of the ACM/IEEE 44th International Conference on Software Engineering: Companion Proceedings}, 2022, pp. 90--94.

\bibitem{taxonomyRealFaults}
N.~Humbatova, G.~Jahangirova, G.~Bavota, V.~Riccio, A.~Stocco, and P.~Tonella, ``Taxonomy of real faults in deep learning systems,'' in \emph{Proceedings of the ACM/IEEE 42nd International Conference on Software Engineering}, ser. ICSE '20.

\bibitem{apriori}
R.~Agrawal and R.~Srikant, ``Fast algorithms for mining association rules in large databases,'' in \emph{Proceedings of the 20th International Conference on Very Large Data Bases}, ser. VLDB '94.\hskip 1em plus 0.5em minus 0.4em\relax San Francisco, CA, USA: Morgan Kaufmann Publishers Inc., 1994, p. 487–499.

\bibitem{motivatingExample1}
\BIBentryALTinterwordspacing
2019, accessed on January 3, 2024. [Online]. Available: \url{https://stackoverflow.com/q/58190114}
\BIBentrySTDinterwordspacing

\bibitem{sequence}
\BIBentryALTinterwordspacing
K.~Team, ``Keras documentation: Python \& numpy utilities.'' [Online]. Available: \url{https://keras.io/2.16/api/utils/python_utils/#sequence-class}
\BIBentrySTDinterwordspacing

\bibitem{motivatingExample2}
\BIBentryALTinterwordspacing
2018, accessed on December 28, 2023. [Online]. Available: \url{https://stackoverflow.com/q/50920908}
\BIBentrySTDinterwordspacing

\bibitem{islam2019fse}
\BIBentryALTinterwordspacing
M.~J. Islam, G.~Nguyen, R.~Pan, and H.~Rajan, ``A comprehensive study on deep learning bug characteristics,'' in \emph{Proceedings of the 2019 27th ACM Joint Meeting on European Software Engineering Conference and Symposium on the Foundations of Software Engineering}, ser. ESEC/FSE 2019.\hskip 1em plus 0.5em minus 0.4em\relax New York, NY, USA: Association for Computing Machinery, 2019, p. 510–520. [Online]. Available: \url{https://doi.org/10.1145/3338906.3338955}
\BIBentrySTDinterwordspacing

\bibitem{croft2023data}
R.~Croft, M.~A. Babar, and M.~M. Kholoosi, ``Data quality for software vulnerability datasets,'' in \emph{2023 IEEE/ACM 45th International Conference on Software Engineering (ICSE)}.\hskip 1em plus 0.5em minus 0.4em\relax IEEE, 2023, pp. 121--133.

\bibitem{jahan2024towards}
S.~Jahan, M.~B. Shah, and M.~M. Rahman, ``Towards understanding the challenges of bug localization in deep learning systems,'' \emph{arXiv preprint arXiv:2402.01021}, 2024.

\bibitem{sanders2010cuda}
J.~Sanders and E.~Kandrot, \emph{CUDA by example: an introduction to general-purpose GPU programming}.\hskip 1em plus 0.5em minus 0.4em\relax Addison-Wesley Professional, 2010.

\bibitem{aviram2012efficient}
A.~Aviram, S.-C. Weng, S.~Hu, and B.~Ford, ``Efficient system-enforced deterministic parallelism,'' \emph{Communications of the ACM}, vol.~55, no.~5, pp. 111--119, 2012.

\bibitem{zhang2018g}
K.~Zhang, B.~He, J.~Hu, Z.~Wang, B.~Hua, J.~Meng, and L.~Yang, ``$\{$G-NET$\}$: Effective $\{$GPU$\}$ sharing in $\{$NFV$\}$ systems,'' in \emph{15th USENIX Symposium on Networked Systems Design and Implementation (NSDI 18)}, 2018, pp. 187--200.

\bibitem{tiwari2015understanding}
D.~Tiwari, S.~Gupta, J.~Rogers, D.~Maxwell, P.~Rech, S.~Vazhkudai, D.~Oliveira, D.~Londo, N.~DeBardeleben, P.~Navaux \emph{et~al.}, ``Understanding gpu errors on large-scale hpc systems and the implications for system design and operation,'' in \emph{2015 IEEE 21st International Symposium on High Performance Computer Architecture (HPCA)}.\hskip 1em plus 0.5em minus 0.4em\relax IEEE, 2015, pp. 331--342.

\bibitem{long2022reporting}
G.~Long and T.~Chen, ``On reporting performance and accuracy bugs for deep learning frameworks: An exploratory study from github,'' in \emph{Proceedings of the 26th International Conference on Evaluation and Assessment in Software Engineering}, 2022, pp. 90--99.

\bibitem{shi2016benchmarking}
S.~Shi, Q.~Wang, P.~Xu, and X.~Chu, ``Benchmarking state-of-the-art deep learning software tools,'' in \emph{2016 7th International Conference on Cloud Computing and Big Data (CCBD)}.\hskip 1em plus 0.5em minus 0.4em\relax IEEE, 2016, pp. 99--104.

\bibitem{SOUsageData}
\BIBentryALTinterwordspacing
S.~Exchange, ``All sites - stack exchange,'' accessed on December 12, 2023. [Online]. Available: \url{https://stackexchange.com/sites?view=list}
\BIBentrySTDinterwordspacing

\bibitem{zhao2021state}
H.~Zhao, Y.~Li, F.~Liu, X.~Xie, and L.~Chen, ``State and tendency: an empirical study of deep learning question\&answer topics on stack overflow,'' \emph{Science China Information Sciences}, vol.~64, pp. 1--23, 2021.

\bibitem{zhang2019empirical}
T.~Zhang, C.~Gao, L.~Ma, M.~Lyu, and M.~Kim, ``An empirical study of common challenges in developing deep learning applications,'' in \emph{2019 IEEE 30th International Symposium on Software Reliability Engineering (ISSRE)}.\hskip 1em plus 0.5em minus 0.4em\relax IEEE, 2019, pp. 104--115.

\bibitem{ponzanelli2014}
L.~Ponzanelli, A.~Mocci, A.~Bacchelli, and M.~Lanza, ``Understanding and classifying the quality of technical forum questions,'' in \emph{2014 14th International Conference on Quality Software}, 2014, pp. 343--352.

\bibitem{StackExchange}
\BIBentryALTinterwordspacing
 [Online]. Available: \url{https://data.stackexchange.com/}
\BIBentrySTDinterwordspacing

\bibitem{replicationPackage}
\BIBentryALTinterwordspacing
M.~Shah, ``mehilshah/bug\_reproducibility\_dl\_bugs,'' accessed on January 3, 2024. [Online]. Available: \url{https://github.com/mehilshah/Bug\_Reproducibility\_DL\_Bugs}
\BIBentrySTDinterwordspacing

\bibitem{cochran1977sampling}
W.~G. Cochran, \emph{Sampling techniques}.\hskip 1em plus 0.5em minus 0.4em\relax john wiley \& sons, 1977.

\bibitem{GeeksforGeeks_ide_dl}
\BIBentryALTinterwordspacing
Jan 2023, accessed on December 25, 2023. [Online]. Available: \url{https://www.geeksforgeeks.org/best-ides-for-machine-learning/}
\BIBentrySTDinterwordspacing

\bibitem{mchugh2012interrater}
M.~L. McHugh, ``Interrater reliability: the kappa statistic,'' \emph{Biochemia medica}, vol.~22, no.~3, pp. 276--282, 2012.

\bibitem{tambon2024silent}
F.~Tambon, A.~Nikanjam, L.~An, F.~Khomh, and G.~Antoniol, ``Silent bugs in deep learning frameworks: an empirical study of keras and tensorflow,'' \emph{Empirical Software Engineering}, vol.~29, no.~1, p.~10, 2024.

\bibitem{pham2020problems}
H.~V. Pham, S.~Qian, J.~Wang, T.~Lutellier, J.~Rosenthal, L.~Tan, Y.~Yu, and N.~Nagappan, ``Problems and opportunities in training deep learning software systems: An analysis of variance,'' in \emph{Proceedings of the 35th IEEE/ACM international conference on automated software engineering}, 2020, pp. 771--783.

\bibitem{alahmari2020challenges}
S.~S. Alahmari, D.~B. Goldgof, P.~R. Mouton, and L.~O. Hall, ``Challenges for the repeatability of deep learning models,'' \emph{IEEE Access}, vol.~8, pp. 211\,860--211\,868, 2020.

\bibitem{gori2023machine}
M.~Gori, A.~Betti, and S.~Melacci, \emph{Machine Learning: A constraint-based approach}.\hskip 1em plus 0.5em minus 0.4em\relax Elsevier, 2023.

\bibitem{liu1999mining}
B.~Liu, W.~Hsu, and Y.~Ma, ``Mining association rules with multiple minimum supports,'' in \emph{Proceedings of the fifth ACM SIGKDD international conference on Knowledge discovery and data mining}, 1999, pp. 337--341.

\bibitem{OpenCV2024}
\BIBentryALTinterwordspacing
Jun 2024. [Online]. Available: \url{https://opencv.org/}
\BIBentrySTDinterwordspacing

\bibitem{PIL}
\BIBentryALTinterwordspacing
Jun 2024. [Online]. Available: \url{https://python-pillow.org/}
\BIBentrySTDinterwordspacing

\bibitem{radford2019language}
A.~Radford, J.~Wu, R.~Child, D.~Luan, D.~Amodei, I.~Sutskever \emph{et~al.}, ``Language models are unsupervised multitask learners,'' \emph{OpenAI blog}, vol.~1, no.~8, p.~9, 2019.

\bibitem{logisticRegression}
\BIBentryALTinterwordspacing
 [Online]. Available: \url{https://machinelearningmastery.com/building-a-logistic-regression-classifier-in-pytorch/}
\BIBentrySTDinterwordspacing

\bibitem{pytorch1.6}
\BIBentryALTinterwordspacing
PyTorch. [Online]. Available: \url{https://pytorch.org/docs/1.6.0/}
\BIBentrySTDinterwordspacing

\bibitem{breck2019data}
E.~Breck, N.~Polyzotis, S.~Roy, S.~Whang, and M.~Zinkevich, ``Data validation for machine learning.'' in \emph{MLSys}, 2019.

\bibitem{soltani2020significance}
M.~Soltani, F.~Hermans, and T.~B{\"a}ck, ``The significance of bug report elements,'' \emph{Empirical Software Engineering}, vol.~25, pp. 5255--5294, 2020.

\bibitem{Chen_Jiang_2022}
\BIBentryALTinterwordspacing
B.~Chen and Z.~M.~J. Jiang, ``\BIBforeignlanguage{en}{A survey of software log instrumentation},'' \emph{\BIBforeignlanguage{en}{ACM Computing Surveys}}, vol.~54, no.~4, p. 1–34, May 2022. [Online]. Available: \url{https://dl.acm.org/doi/10.1145/3448976}
\BIBentrySTDinterwordspacing

\bibitem{talwar2020evaluating}
D.~Talwar, S.~Guruswamy, N.~Ravipati, and M.~Eirinaki, ``Evaluating validity of synthetic data in perception tasks for autonomous vehicles,'' in \emph{2020 IEEE International Conference On Artificial Intelligence Testing (AITest)}.\hskip 1em plus 0.5em minus 0.4em\relax IEEE, 2020, pp. 73--80.

\bibitem{grosse2017lecture}
R.~Grosse, ``Lecture 15: Exploding and vanishing gradients,'' \emph{University of Toronto Computer Science}, 2017.

\bibitem{Tishby_Zaslavsky_2015}
\BIBentryALTinterwordspacing
N.~Tishby and N.~Zaslavsky, ``Deep learning and the information bottleneck principle,'' in \emph{2015 IEEE Information Theory Workshop (ITW)}, Apr. 2015, p. 1–5. [Online]. Available: \url{https://ieeexplore.ieee.org/abstract/document/7133169}
\BIBentrySTDinterwordspacing

\bibitem{yang2022comprehensive}
Y.~Yang, T.~He, Z.~Xia, and Y.~Feng, ``A comprehensive empirical study on bug characteristics of deep learning frameworks,'' \emph{Information and Software Technology}, vol. 151, p. 107004, 2022.

\bibitem{nelder1972generalized}
J.~A. Nelder and R.~W. Wedderburn, ``Generalized linear models,'' \emph{Journal of the Royal Statistical Society Series A: Statistics in Society}, vol. 135, no.~3, pp. 370--384, 1972.

\bibitem{ceccato2014family}
M.~Ceccato, M.~Di~Penta, P.~Falcarin, F.~Ricca, M.~Torchiano, and P.~Tonella, ``A family of experiments to assess the effectiveness and efficiency of source code obfuscation techniques,'' \emph{Empirical Software Engineering}, vol.~19, pp. 1040--1074, 2014.

\bibitem{Opinio}
\BIBentryALTinterwordspacing
 [Online]. Available: \url{https://surveys.dal.ca/opinio/admin/folder.do}
\BIBentrySTDinterwordspacing

\bibitem{mondal2022reproducibility}
S.~Mondal, M.~M. Rahman, C.~K. Roy, and K.~Schneider, ``The reproducibility of programming-related issues in stack overflow questions,'' \emph{Empirical Software Engineering}, vol.~27, no.~3, p.~62, 2022.

\bibitem{mondalreproducibility}
S.~Mondal and B.~Roy, ``Reproducibility of issues reported in stack overflow questions: Challenges, impact \& estimation,'' \emph{Impact \& Estimation}.

\bibitem{rahman2022works}
M.~M. Rahman, F.~Khomh, and M.~Castelluccio, ``Works for me! cannot reproduce--a large scale empirical study of non-reproducible bugs,'' \emph{Empirical Software Engineering}, vol.~27, no.~5, p. 111, 2022.

\bibitem{tfProgramBugs}
\BIBentryALTinterwordspacing
Y.~Zhang, Y.~Chen, S.-C. Cheung, Y.~Xiong, and L.~Zhang, ``An empirical study on tensorflow program bugs,'' in \emph{Proceedings of the 27th ACM SIGSOFT International Symposium on Software Testing and Analysis}, ser. ISSTA 2018.\hskip 1em plus 0.5em minus 0.4em\relax New York, NY, USA: Association for Computing Machinery, 2018, p. 129–140. [Online]. Available: \url{https://doi.org/10.1145/3213846.3213866}
\BIBentrySTDinterwordspacing

\bibitem{tarek2022icsme}
T.~Makkouk, D.~J. Kim, and T.-H.~P. Chen, ``An empirical study on performance bugs in deep learning frameworks,'' in \emph{2022 IEEE International Conference on Software Maintenance and Evolution (ICSME)}, 2022, pp. 35--46.

\bibitem{nerdbug}
\BIBentryALTinterwordspacing
F.~Jafarinejad, K.~Narasimhan, and M.~Mezini, ``Nerdbug: Automated bug detection in neural networks,'' in \emph{Proceedings of the 1st ACM International Workshop on AI and Software Testing/Analysis}, ser. AISTA 2021.\hskip 1em plus 0.5em minus 0.4em\relax New York, NY, USA: Association for Computing Machinery, 2021, p. 13–16. [Online]. Available: \url{https://doi.org/10.1145/3464968.3468409}
\BIBentrySTDinterwordspacing

\bibitem{yan2021exposing}
M.~Yan, J.~Chen, X.~Zhang, L.~Tan, G.~Wang, and Z.~Wang, ``Exposing numerical bugs in deep learning via gradient back-propagation,'' in \emph{Proceedings of the 29th ACM Joint Meeting on European Software Engineering Conference and Symposium on the Foundations of Software Engineering}, 2021, pp. 627--638.

\bibitem{zhang2020detecting}
Y.~Zhang, L.~Ren, L.~Chen, Y.~Xiong, S.-C. Cheung, and T.~Xie, ``Detecting numerical bugs in neural network architectures,'' in \emph{Proceedings of the 28th ACM Joint Meeting on European Software Engineering Conference and Symposium on the Foundations of Software Engineering}, 2020, pp. 826--837.

\bibitem{trainingReproducibleDL}
\BIBentryALTinterwordspacing
B.~Chen, M.~Wen, Y.~Shi, D.~Lin, G.~K. Rajbahadur, and Z.~M.~J. Jiang, ``Towards training reproducible deep learning models,'' in \emph{Proceedings of the 44th International Conference on Software Engineering}, ser. ICSE '22.\hskip 1em plus 0.5em minus 0.4em\relax New York, NY, USA: Association for Computing Machinery, 2022, p. 2202–2214. [Online]. Available: \url{https://doi.org/10.1145/3510003.3510163}
\BIBentrySTDinterwordspacing

\bibitem{reproducibilityDLTSE}
\BIBentryALTinterwordspacing
C.~Liu, C.~Gao, X.~Xia, D.~Lo, J.~Grundy, and X.~Yang, ``On the reproducibility and replicability of deep learning in software engineering,'' \emph{ACM Trans. Softw. Eng. Methodol.}, vol.~31, no.~1, oct 2021. [Online]. Available: \url{https://doi.org/10.1145/3477535}
\BIBentrySTDinterwordspacing

\bibitem{mondal2024can}
S.~Mondal, M.~M. Rahman, and C.~K. Roy, ``Can we identify stack overflow questions requiring code snippets? investigating the cause \& effect of missing code snippets,'' \emph{arXiv preprint arXiv:2402.04575}, 2024.

\bibitem{reproducibleBugReports}
M.~White, M.~Linares-Vásquez, P.~Johnson, C.~Bernal-Cárdenas, and D.~Poshyvanyk, ``Generating reproducible and replayable bug reports from android application crashes,'' in \emph{2015 IEEE 23rd International Conference on Program Comprehension}, 2015, pp. 48--59.

\bibitem{rutherford2011anova}
A.~Rutherford, \emph{ANOVA and ANCOVA: a GLM approach}.\hskip 1em plus 0.5em minus 0.4em\relax John Wiley \& Sons, 2011.

\end{thebibliography}

\newpage
\appendix
\section*{Appendix A: Flowchart for Manual Analysis}
\begin{figure}[h]
    \centering
    \includegraphics[width=0.7\textheight, angle = 90]{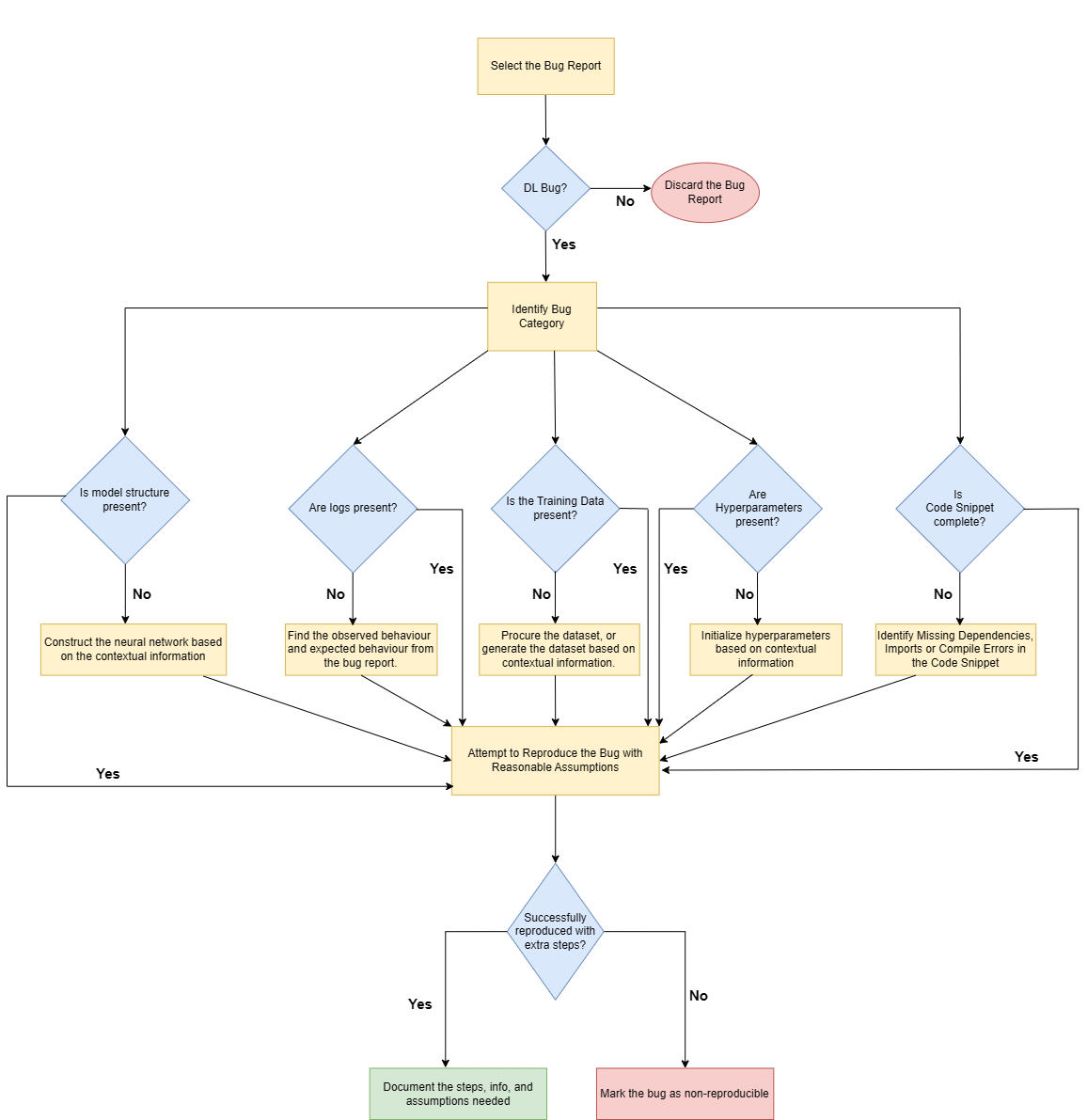}
    \caption{Manual Analysis of Stack Overflow Posts}
    \label{fig:manualAnalysisSOPosts}
\end{figure}

The flowchart in Fig.~\ref{fig:manualAnalysisSOPosts} illustrates a systematic process for analyzing Stack Overflow posts related to deep learning bugs. It begins by determining if the post discusses a deep-learning bug. If it does, the bug category is identified based on the symptoms, potential root causes, post tags and explanation by the user. The flowchart then guides the user through a series of checks to assess the reproducibility of the bug, considering factors such as the completeness of the code snippet, availability of necessary data, presence of relevant logs or error messages, specification of hyperparameters, and clarity of the model structure. We then try to reproduce the bug. If the reproduction is successful with extra steps, the steps, information, and assumptions needed are documented, or else the bug is marked as non-reproducible.

\appendix
\section*{Appendix B: Improving the Reproducibility of Deep Learning Bugs using LLMs}
In this section, we further aim to verify the practical utility of our findings in bug reproduction. To achieve this, we leverage the capabilities of a large language model (LLM) and offer more actionable recommendations to developers.

\begin{figure}[h]
    \centering
    \includegraphics[width=0.90\linewidth]{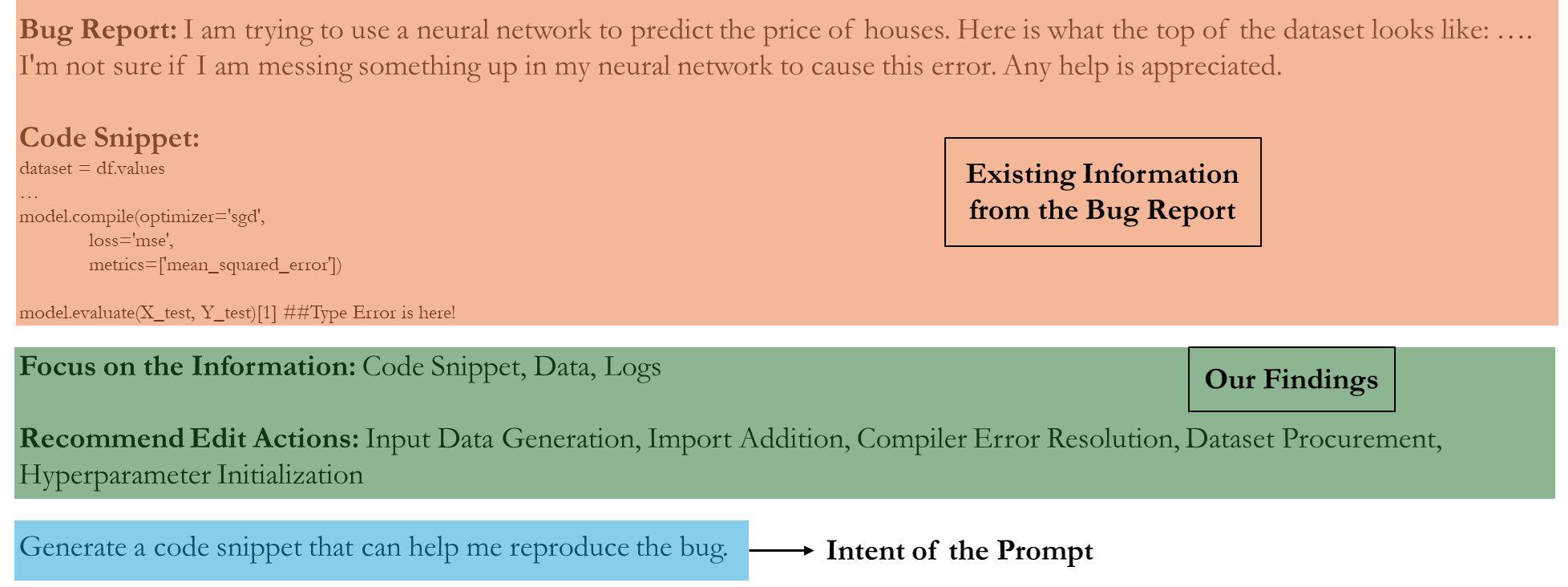}
    \caption{Prompt for LLaMA3}
    \label{fig:prompt}
\end{figure}

We conduct experiments using the LLaMA3, a 70B parameter LLM, to evaluate its effectiveness in assisting developers with the reproduction of deep learning bugs. To conduct these experiments, we construct a dataset of 40 bugs. We used stratified random sampling and collect 40 of our manually reproduced bugs, with the following distribution: 10 Training, 10 API, 10 Tensor and 10 Model Bugs. For each of these four bug categories, we ensure that the dataset contains an equal number of easy and difficult bugs, determined by the number of edit actions required to reproduce the bug. Specifically, within each bug category, there are 5 bugs that require a relatively small number of edit actions (classified as easy), and 5 bugs that require a larger number of edit actions (classified as difficult). The prompt for the LLaMA3 model contained four key inputs: the bug report describing the issue, relevant code snippets providing context, the instruction to generate a code snippet that could reproduce the reported bug, and, crucially, the edit actions and component information derived from our findings. Fig.~\ref{fig:prompt} demonstrates an example prompt for our experiments. Once the model generates the code snippets, we check their ability to reproduce the bugs through manual execution and comparison with the erroneous behaviour outlined in the bug reports. Furthermore, we compare the generated code with the ground truth code snippets curated during our manual bug reproduction process.

The results of our experiments are as follows:
\begin{itemize}
\item \textbf{Performance without Augmentation:} When the LLM was provided with only the bug report and code snippet, its suggestions were less useful. Its generated code snippets lack correctness, despite being compilable and executable. The absence of guidance from our findings diminishes the LLM's performance, with the model generating reproducible code snippets for only 25 out of the 40 bugs (\textbf{62.5\%}).
\item \textbf{Performance with Augmentation using Hints:} When guided by our edit actions and component information, the LLM generates reproducible code snippets for 33 out of the 40 bugs (\textbf{82.5\%}). Analysis of the generated code snippets reveals that the model effectively understands the bug, uses appropriate edit actions, and generates complete, compilable, and executable code snippets. Additionally, in some scenarios, the model provides a textual description of the circumstances under which the bug might be reproducible, offering valuable extra information to developers. Code snippets generated by Llama3, which demonstrate the model's improved understanding after augmentation using hints, are also available in the replication package~\cite{replicationPackage}.
\end{itemize}

\textbf{Manual Analysis of Generated Code Snippets:} We perform manual analysis of the generated code snippets to understand the behaviour of LLMs. From the manual analysis of the generated code snippets, we observed that the LLM performs relatively well in generating code snippets for Training and Model bugs, even without augmentation. This behaviour could be explained by the high prevalence of these bugs in real-life scenarios~\cite{taxonomyRealFaults}. On the other hand, the LLM faces more challenges in generating the code snippets for reproducing the Tensor and API bugs without guidance. Tensor bugs usually involve complex manipulations of tensor shapes and dimensions, which might not be fully understood by LLMs. Similarly, API bugs are sensitive to constant updates to the API documentation. The frequent changes to the API documentation make it harder for LLMs to stay up-to-date and generate accurate code snippets to reproduce these bugs. However, when provided with our guidance, the reproducibility of API bugs goes up from 56\% to 76\%, and Tensor bugs reproducibility goes up from 58\% to 75\%. This demonstrates the utility of our findings in improving the ability of language models to generate code snippets for bugs that are relatively difficult to reproduce.

Our experiments demonstrate that augmenting the LLM with our suggested edit actions and component information enhances its effectiveness in reproducing deep learning bugs by 20\%. This improvement underscores the potential of integrating our findings with LLMs to provide developers with more actionable guidance for reproducing deep learning bugs.

\subsection*{Statistical Significance Tests for our Experiments}
Since the recommendations of LLMs are stochastic by design, different runs with the same prompt can produce different reproducible snippets. This introduces an additional source of variability which may invalidate our previous observations. To address this issue, we repeat the code generation task for each bug five times and record their rates of success in bug reproduction. To assess the statistical significance of the LLM results, we employ the RM-ANOVA test~\cite{rutherford2011anova}. RM-ANOVA is suitable for this scenario because we have a within-subjects design with one independent variable (augmentation with hints or not) and one dependent variable (bug reproduction rate) measured repeatedly across trials. The test yields a significant p-value of $5.8 \times 10^{-5}$, demonstrating the positive impact of hints on bug reproduction abilities of large language models. Additionally, we have used the partial-eta squared ($\eta_p^2$) metric for the effect size, as it is a common measure in ANOVA representing the proportion of total variance attributable to a factor adjusted for other factors in the model. The $\eta_p^2$ for the Guidance factor (hint augmentation) is 0.188324, which is considered a large effect size. This substantial effect size, coupled with the significant p-value, provides strong evidence for both the statistical and practical significance of using hints to improve LLM performance in bug reproduction tasks.
\end{document}